\documentclass[12pt]{article}
\usepackage{jheparxiv}

\usepackage{graphicx}
\usepackage{slashed}

\usepackage{tikzit}
\tikzstyle{new style 0}=[fill=black, draw=black, shape=circle, radius=0.1]
\usetikzlibrary{shapes.geometric, arrows}
\usepackage{slashed}
\usepackage{nicefrac}
\usepackage{hhline}
\usepackage{braket}

\usepackage{amsmath}
\usepackage{color}

\usepackage[english]{babel}
\usepackage{times}
\usepackage[T1]{fontenc}
\usepackage{latexsym}
\usepackage{mathrsfs}
\usepackage{amssymb}
\newcommand{\half}{\frac{\scriptstyle 1}{\scriptstyle 2}}

\newcommand{\C}{\mathbb{C}}

\newcommand{\HH}{\mathbb{H}}
\newcommand{\CP}{\mathbb{CP}}
\newcommand{\RP}{\mathbb{RP}}
\newcommand{\PT}{\mathbb{PT}}
\newcommand{\PS}{\mathbb{PS}}
\newcommand{\R}{\mathbb{R}}
\renewcommand{\P}{\mathbb{P}} 
\newcommand{\bbS}{\mathbb{S}}
\newcommand{\bbD}{\mathbb{D}}

\newcommand{\cH}{\mathcal{H}}
\newcommand{\scri}{\mathscr{I}}

\newcommand{\M}{\mathbb{M}}
\newcommand{\cM}{M}
\newcommand{\CM}{\mathcal{M}}
\newcommand{\cD}{\mathcal{D}}

\newcommand{\cN}{\mathcal{N}}

\newcommand{\N}{\mathbb{N}}
\newcommand{\T}{\mathbb{T}}
\newcommand{\Z}{\mathbb{Z}}
\newcommand{\p}{\partial}
\newcommand{\dbar}{\bar\partial}

\newcommand{\e}{\mathrm{e}}

\newcommand{\D}{\mathrm{D}}

\newcommand{\cA}{\mathcal{A}}

\newcommand{\cO}{\mathcal{O}}

\renewcommand{\P}{\mathbb{P}}

\newcommand{\diag}{\, \mathrm{diag}}
\newcommand{\tr}{\, \mathrm{tr}}

\newcommand{\bx}{{{\mathbf{x}}}}
\newcommand{\by}{{{\mathbf{y}}}}

\newcommand{\bigma}{\boldsymbol{\sigma}}
\newcommand{\be}{\begin{equation}\label}
\newcommand{\ee}{\end{equation}}
\newcommand{\bea}{\begin{eqnarray}\label}
\newcommand{\eea}{\end{eqnarray}}

\newtheorem{defn}{Definition}
\newtheorem{thm}{Theorem}

\newtheorem{propn}{Proposition}

\title{ Gravity from holomorphic discs and celestial   $Lw_{1+\infty}$ symmetries }

\author{ Lionel Mason} \affiliation{The IHES, 35 Rte du Chartres, Bures Sur Yvette,  France 91440,
\footnote{Also Laboratoire de Physiques ENS, Rue Homond, Paris 7005, on leave from The
  Mathematical Institute, AWB ROQ, Woodstock Rd,  Oxford OX1 6GG }
  \\ \texttt{lmason@maths.ox.ac.uk}}

\date{\today}  
\begin{document}

\abstract{
In split or Kleinian signature, twistor constructions parametrize  solutions to both gauge and gravity self-duality (SD)  equation from  twistor data that can be expressed in terms of free smooth data  without gauge freedom. Here the corresponding constructions are given  for asymptotically flat SD gravity providing a fully nonlinear encoding of the asymptotic gravitational  data in terms of  a  real homogeneous generating function $h$ on the real twistor space.  Geometrically $h$ determines a nonlinear deformation of the location of the real twistor space $\RP^3$ inside the complex twistor space $\CP^3$.   

This presentation gives an optimal presentation of Strominger's  recently discovered $Lw_{1+\infty}$ celestial symmetries.  These,  when real,  act locally as passive Poisson diffeomorphisms  on the real twistor space.  However, when imaginary,  such Poisson transformations are active symmetries, and  generate changes to  the gravitational field by deforming the location of the real slice of the twistor space.  

Gravity amplitudes for the full, non-self-dual Einstein gravity, arise as correlators of a chiral  twistor  sigma model.  This is reformulated for split signature as a theory of holomorphic discs in twistor space whose boundaries lie on the deformed real slice  determined by  $h$.  Real $Lw_{1+\infty}$ symmetries act oas gauge symmetries, but imaginary generators yield graviton vertex operators that generate gravitons in the perturbative expansion. 

A generating function for the all plus 1-loop amplitude, an analogous framework for Yang-Mills, possible interpretations in Lorentz signature and similar open string formulations of twistor and ambitwistor strings in 4d in split signature, are briefly discussed.
}

\maketitle

%\tableofcontents

\section{Introduction}

Celestial Holography seeks to encode  4d gravity in an asymptotically flat space-time from a boundary theory that is defined at the null infinity, 
%soft OPEs of gravity S-matrix  at 
$\scri$, that arises when the cosmological constant vanishes \cite{Strominger:2017zoo,McLoughlin:2022ljp}. 
Similarly, in the 1970's Newman  tried to rebuild space-time from the  `cuts' of $\scri$ formed by the intersection of light cones of points in the interior of space-time with $\scri$, \cite{Newman:1976gc}. His `good cut' equation   instead yielded $\cH$-space, a complex self-dual space-time built from the self-dual part of the characteristic data at $\scri$.  Penrose subsequently re-interpreted this as a  twistor construction \cite{Penrose:1976js} by first introducing an `asymptotic twistor space' \cite{Ko:1977gw,Hansen:1978jz} defined from data at $\scri$. More recently, it has been possible to use this framework to construct sigma models and strings whose target spaces are these asymptotic twistor spaces or their cotangent bundles (ambitwistor spaces)   \cite{Adamo:2014yya, Geyer:2014lca,Adamo:2019ipt,Adamo:2021bej,Adamo:2022mev}.  These then yield explicit formulae for complete tree-level S-matrices for gauge and gravity theories based on twistor-string ideas \cite{Witten:2003nn, Berkovits:2004hg, Skinner:2013xp} and ambitwistor-string ideas \cite{Mason:2013sva, Geyer:2014fka,Geyer:2022cey} reformulated at null infinity using these asymptotic twistor and ambitwistor  spaces as targets.  These theories lead to easy computations \cite{Adamo:2014yya, Geyer:2014lca,Adamo:2019ipt,Adamo:2021bej,Adamo:2022mev} of the soft limits  \cite{McLoughlin:2022ljp} on which much of celestial holography is based: such soft limits can be taken at the level of the vertex operators of these models  \cite{Adamo:2014yya, Geyer:2014lca,Adamo:2019ipt,Adamo:2021lrv, Adamo:2021tba}, for a   review   see \cite{Geyer:2022cey}. Particularly in the form of \cite{Adamo:2021bej,Adamo:2022mev}, these formulae are based on rational curves that can be thought of as extending Newman's construction to provide the formulae for the full light-cone cuts of infinity in the form of an expansion around the self-dual sector in such a way as to provide also formulae for the tree-level gravitational S-matrix.   These models therefore provide reformulations at $\scri$ of gauge and gravity theories. 
The twisted celestial holography programme of Costello and Paquette  \cite{Costello:2022wso,Costello:2022jpg} also lives in twistor space, and is perhaps most easily understood via the same mechanism.

This paper reformulates this twistor theory in split signature, i.e.\ with metric that has diagonal entries $(1,1,-1,-1)$, which is sometimes also known as Kleinian signature. Split signature has already been explored in the context of the linear Penrose transform and its applications  \cite{Woodhouse:1992,  Sparling:1998, Witten:2003nn,Mason:2009sa,Arkani-Hamed:2009hub,Arkani-Hamed:2012zlh}, and also in various nonlinear regimes in \cite{Mason:1996in,Lebrun:2002,Mason:2005qu,Lebrun:2007,Lebrun:2009,Lebrun:2010}.
More recently it has been 
used in the context of celestial holography \cite{Atanasov:2021oyu} to understand in particular the recently discovered $Lw_{1+\infty}$ symmetry \cite{Guevara:2021abz, Strominger:2021lvk} see also \cite{Guevara:2021yud,Brown:2022miw} for other recent applications of twistors in split signature.  In \cite{Adamo:2021lrv} these symmetries were interpreted as the natural structure-preserving diffeomorphisms on twistor space constructed in Lorentz signature.  However,   the correspondence was obscured by the need to perform a translation from a  \v Cech to a  Dolbeault presentation.  This was analogous to the shadow and light-ray transforms used in \cite{ Strominger:2021lvk,Sharma:2021gcz, Brown:2022miw, Jorge-Diaz:2022dmy}.  In split signature we  will see that the twistor theory automatically provides the framework in which the $Lw_{1+\infty}$ appears as the natural structure preserving diffeomorphisms of the twistor space more directly than was possible in \cite{Adamo:2021lrv}. 

The use of split signature perhaps requires some comment.  The  analytic continuation  of scattering amplitudes and correlation functions has always been a powerful tool, although mostly to Euclidean signature  to renders QFT quantities  more regular.  However, there is much physics in their  singularity structure  and their geometry can be made explicit in split signature particularly at tree level and for integrands.  In four dimensions, such a continuation was used  to uncover the twistor support of amplitudes in the original twistor-string \cite{Witten:2003nn}, and in BCFW recursion in \cite{ Mason:2009sa,Arkani-Hamed:2009hub} and subsequently in the positive Grassmannian \cite{Arkani-Hamed:2012zlh} and the amplituhedron \cite{Arkani-Hamed:2013jha}.  In this article we will see that it also uncovers quite explicitly the underlying geometry of the $Lw_{1+\infty}$ symmetry discovered recently by Strominger and co-workers \cite{Guevara:2021tvr, Guevara:2021abz,Strominger:2021lvk}.

The utility of split signature arises because the Penrose transform  usually represents space-time massless fields in terms of cohomology classes on twistor space.  These   are subject to cohomological  gauge freedoms that can obscure the underlying geometry.  However,  it was observed early on that in split signature, the Penrose transform becomes the X-ray transform of Fritz-John \cite{John:1938}.  This gives an essentially  one to one map between functions on the totally real twistor space, $\RP^3$ and solutions to zero rest mass equations on split signature space-time. The transform integrates a smooth function on the real twistor space $\RP^3$  along straight lines.  Via the Klein correspondence, this space of lines is a 4-dimensional real quadric  in $\RP^5$ that can be understood   as the conformal compactification of split signature Minkowski space $\M^{2,2}$ \cite{Gelfand:2003}.
% (or, as the integration requires a choice of orientation, its double cover as we will see later). 
%This correspondence rigidifies the Penrose transform, removing the gauge freedom in the twistor representations  as it is precisely 1:1 between global  solutions to the wave equation and smooth functions of appropriate weights on $\RP^3$.
This has been explored using the complex twistor geometry  for the  linear transform in \cite{ Bailey:1999,Bailey:2002}  and in particular for amplitudes via Witten's half-Fourier transform from momentum space  \cite{Witten:2003nn} in his study of the twistor support of gauge theory amplitudes on rational curves in twistor space arising from  twistor-string theory.  The nonlinear analogues are quite nontrivial and have been part of an ongoing investigation  with Claude Lebrun in \cite{Lebrun:2002, Lebrun:2007, Lebrun:2009, Lebrun:2010} for the nonlinear constructions relating to conformal and projective geometry, and  in \cite{Mason:1996in, Mason:2005qu} for self-dual Yang-Mills.  These works did not explicitly work  out the details for self-dual Einstein gravity and the first  focus of this paper is to give a similar treatment for SD Einstein vacuum spaces with vanishing cosmological constant.

To summarize this part, in split signature there are real twistors that  correspond to real self-dual totally  null 2-surfaces in space-time.  In flat space  these real twistors make up the real slice $\PT_\R=\RP^3$, as fixed by the standard complex conjugation inside complex projective twistor space $\PT=\CP^3$.  
In the global split signature construction of \cite{Lebrun:2007}, it is shown that under suitable global assumptions, the complex holomorphic twistor space can still be reconstructed as $\CP^3$, as in the standard flat model, but the location of the real slice $\PT_\R$ is deformed essentially arbitrarily away from the standard embedding of $\RP^3$.  The self-dual space-time is then reconstructed as the moduli space of degree-1 holomorphic discs whose boundary lies in this deformed $\PT_\R$.  A particular feature is that the space of such discs has topology $S^2\times S^2$ whereas $\M^{2,2}=S^2\times S^2/\Z_2$, because in the flat case two discs make up each real Riemann sphere  that is sent to itself under the complex conjugation that  fixes  $\RP^3$.  In the flat case the $\Z_2$ action on $S^2\times S^2$ whose quotient gives $\M^{2,2}$ is 
the joint antipodal map on the $S^2$ factors.  In the nonlinear case, once  the location of $\PT_\R$ has been deformed, global solutions on $S^2\times S^2$ are  found at least for small deformations, but only the conformally flat metric is $\Z^2$-invariant descending to $S^2\times S^2/\Z^2$.

This paper incorporates the Einstein equations with vanishing cosmological constant into the construction of \cite{Lebrun:2007} described above.  In order to impose  this Einstein condition, we must endows the complex twistor space with a global but degenerate  holomorphic Poisson structure that is real on $\RP^3$.  Introduce   $Z=(\lambda_\alpha,\mu^{\dot\alpha})=U+iV$ as homogeneous coordinates on $\PT$, where $\alpha=0,1$,  and $\dot\alpha=\dot 0,\dot1$ are standard two-component spinor indices for the Lorentz group in 4d and $U$ and $V$ represent the real and imaginary parts.  Then the Poisson structure is given explicitly by    
\begin{equation}
 \{g_1,g_2\}=\varepsilon^{\dot\alpha\dot\beta}\frac{\p g_1}{\p u^{\dot\alpha}}\frac{\p g_2}{\p u^{\dot\beta}}\, .
\end{equation}
Given this,   the gravitational data is encoded into  a real homogeneous function $h(U)$ of weight 2 on $\RP^3$.  It has the interpretation as the generating function that shifts the standard $\RP^3$ into its deformed position $\PT_\R$ on which the restriction of the Poisson structure remains real. Explicitly we have
\begin{equation}
\PT_\R=\left\{[Z=U+iV]\in \PT| V= \{h,U\}\right\}\, .
\end{equation}
This  gives a clean geometric realisation of the recently discovered $Lw_{1+\infty}$ symmetries and its action on the gravitational data.  
The $Lw_{1+\infty}$ in this signature becomes  the real Poisson diffeomorphisms $\mathcal P$  of the real twistor space $\PT_\R$.  This acts naturally and geometrically on this space of gravitational data as a passive diffeomorphism freedom.  In the twistor representation, the data represents points in the quotient $\mathcal P_\C/\mathcal{P}$ of the complex Poisson diffeomorphism semi-group $\mathcal P_\C$,   by the real Poisson diffeomorphisms $\mathcal{P}$.  Here $\mathcal P_\C$ is really a semi-group defined as a higher-dimensional analogue of Segal's semi-group of annuli that complexify the diffeomorphism group of the circle as in  \cite{Segal:2002ei}. Our generating functions $h(U)$ parametrize a natural choice of imaginary slice of $\mathcal{P}_\C$. 

In order to connect to celestial formulations of gravity, we explain how the twistor gravitational data can be constructed directly from  the asymptotic gravitational data, the \emph{asymptotic shear} $\bigma$,  at null infinity $\scri$.   This  uses 
the fact that $\bigma$ determines a  projective structure on each $\beta$-plane (ASD totally null 2-plane) that lies on $\scri$; its geodesics are the null geodesics on $\scri$ that lie in the $\beta$-planes.  Because they bound $\alpha$-surfaces in the interior space-time, they must be closed circles.  Such projective structures were studied in \cite{Lebrun:2002} via a two dimensional analogue of the twistor correspondence described above. This allows us to establish a fully nonlinear correpsondence between the twistor data $h(U)$  and the asymptotic shears $\bigma$ at $\scri$; this was expressed via a \v Cech-Dolbeault correspondence in \cite{Adamo:2021lrv} and generalizes the linear-theory light-ray transform used by \cite{Strominger:2021lvk, Sharma:2021gcz} to establish the correspondence between asymptotic shears and generators of $Lw_{1+\infty}$.

% we can adapt twistor and  ambitwistor string models to this geometry.  They are usually formulated as closed string models with worldsheets of all topological degree.  At  degree-1 they provide the Riemann spheres that play a central role in Penrose's original nonlinear graviton construction for self-dual Einstein metrics \cite{Penrose:1976jq}, and Ward's construction for self-dual Yang-Mi lls fields \cite{Ward:1977ta}.  These are fully nonlinear constructions: in the Penrose nonlinear graviton construction, the self-dual Einstein manifold is reconstructed as the moduli space of holomorphic degree-1 curves inside a deformed twistor space.  In the series of articles with Claude LeBrun, 
%the global twistor constructions in split signature are based instead on holomorphic discs with boundary conditions. 

To make contact with amplitudes, soft theorems etc., we adapt the worldsheet models that give rise to  twistorial realizations of the gravity S-matrix in twistor space to this geometry focussing on the chiral sigma model of \cite{Adamo:2021bej}.
\footnote{The $\cN=8$ twistor string of \cite{Skinner:2013xp} could have been similarly adapted.}  
This was  formulated there as a closed string model, but the twistor geometry in split signature  naturally lends itself to an open-string  formulation.  The strings are required to be holomorphic maps from a Riemann surface with boundary, a holomorphic disk $\D$  at tree-level, to the complex twistor space with boundary lying in $\PT_\R$.   We focus on the  twistor chiral sigma models whose classical correlation functions  compute the full tree-level gravity S-matrix as in \cite{Adamo:2021bej}.
%,  although we could as well have used the open string analogues of  the Skinner model \cite{Skinner:2013xp} whose quantum correlators give the full (i.e., not self-dual)  gravity S-matrix.  
To compute amplitudes with $k$ negative helicity particles, the open chiral sigma model is a theory of maps $Z:\D\rightarrow \PT$ of degree $k-1$ where $\D$ is a holomorphic disc, represented as the upper-half-plane, with coordinate $\sigma$ with action
\begin{equation}
S[Z]= \Im \int_\D \varepsilon_{\dot\alpha\dot\beta}\mu^{\dot\alpha}\dbar \mu^{\dot\beta} d\sigma +\oint_{\p\D} h(U) d\sigma \, .
\end{equation} 
The boundary term gives rise to the required boundary condition that $\p\D\subset \PT_\R$ rather than $\RP^3$.  For an amplitude with $k$ ASD  particles on the SD background determined by $h$, we require that at  $k$ points $\sigma_i\in \p\D$, $Z(\sigma)$ is required to go through the given twistors $Z_i$ at $\sigma_i$.  The tree-level amplitude for $k$ ASD particles on the fully nonlinear background determined by $h$ is  built from the corresponding on-shell action $S^{os}[Z_i\sigma_i,h]$  for the unique degree $k$ solution.  This must then be integrated against the wave functions for the ASD particles inserted at the $\sigma_i$ and $Z_i$ together with a reduced determinant. See \eqref{amp-background} for full details.  To prove this, one can expand the on-shell action $S^{os}[Z_i,\sigma_i,h]$
perturbatively in $h$ and one then obtains a version of the Cachazo-Skinner formula \cite{Cachazo:2012kg, Cachazo:2012pz} as in \cite{Adamo:2021bej}.
One can see from this presentation that the formula is invariant under real $Lw_{1+\infty}$ motions at all MHV degrees, whereas  the SD gravity vertex operator gives the action on the Sigma model of an imaginary  generator of  $Lw_{1+\infty}$.

This article reviews much of the background in order to be  self-contained.  Thus \S \ref{flat-geom} reviews the global geometry of compactified conformally flat space-time in split signature, how null infinity  $\scri$ sits inside it, and their correspondence with flat twistor space making contact with \cite{Atanasov:2021oyu} but with different coordinates adapted to the twistor correspondence.  The next section \S \ref{gen-functions} starts in \S\ref{Rev-confSD}  by reviewing the conformally self-dual case discussed in \cite{Lebrun:2007}.  It goes on to specialise to the Ricci-flat case, introducing the generating function $h$ in \S\ref{gen-fns} and discussing their connections with the $Lw_{1+\infty}$ symmetry in \S\ref{Lw1+infty}. The next section \S\ref{transform-scri} explains the nonlinear correspondence between the twistor data $h(U)$ and the asymptotic shear $\bigma$ on $\scri$ for this class of space-times. Section \ref{Open-model} introduces the open chiral sigma model and presents the amplitude formulae for the tree-level gravitational S-matrix.
The last section \S\ref{discussion} gives a brief summary and discussion.  This includes some remarks on generating formulae for the all-plus 1-loop amplitude, the translation of the results into Lorentz signature, the corresponding open twistor and ambitwistor models, and the Yang-Mills analogues of the ideas in this paper.  Appendix \ref{int-formulae} reviews the linear theory integral formulae that arise in the linearized limit of the transforms described in this paper and their relationship with others such as the Fourier transform, half-Fourier transform and the Kirchoff-d'Adhemar formulae from asymptotic data. In particular the map between the asymptotic shear and $h(U)$ in linear theory can be realized as a Radon transform from $h$ to give $\bigma$.  The second appendix \ref{taub-nut} describes how the Gibbons-Hawking ansatze adapts to split signature in the framework of this paper to give many regular solutions, and also the (singular) SD taub-NUT metric in split  signature, as described in \cite{Crawley:2021auj}.

\section{The twistor geometry and $\scri$ and in split signature} \label{flat-geom}
In this section we review the global split signature geometry of conformal compactifications of Minkowski space; see also \cite{Mason:2005qu, Atanasov:2021oyu} for a different but closely related discussion.  We go on to describe  its  correspondence with twistor space and relegate to appendix \S\ref{int-formulae} the  various integral formulae for the wave equations in terms of the Fourier transform, the X-ray transform to twistor space and the Kirchoff-d'Adhemar formulae from $\scri$.

\subsection{Compactified Minkowski space and $\scri$}
Split signature flat Minkowski space  $\R^{2+2}$ has   conformal compactification, denoted $\M^{2+2}$, of  topology $(S^2 \times S^2)/\Z_2$. In this signature, the conformal group is $SO(3,3)/\Z_2$ acting naturally on the `embedding space' representation 
\begin{equation}
 \M^{2+2}:=\{X\in \R^{3+3}|X\cdot X=0\}/\sim\, , \qquad X\sim a X\, .
\end{equation}
The  `embedding space' coordinates can be  divided  into a pair of three-vectors $X:=(\bx,\by)\in \R^3\times \R^3$ so that  $X\cdot X:=|\bx|^2-|\by|^2$.  If we fix the  rescaling freedom by $|\bx|=|\by|=1$, we see that the double cover is $S^2\times S^2$.  This gives the choice of conformal  metric 
\begin{equation}
ds^2_{S^2\times S^2}=( ds^2_{S^2_\bx} - ds^2_{S^2_\by})\, .
\end{equation}
However, to obtain $\M^{2+2}$ we must quotient also by the $\Z_2$ that acts by 
\begin{equation}
(\bx,\by)\rightarrow (-\bx,-\by)\label{Z2}
\end{equation}
without fixed points thereby obtaining $(S^2\times S^2)/\Z_2$ as claimed.

To make the conformal flatness more explicit, these coordinates can be related to standard affine coordinates by introducing the complex stereographic coordinates
\begin{equation}
z=: \frac{x_1+ix_2}{x_3-y_3}\, , \qquad w :=\frac{y_1+iy_2}{x_3-y_3}\, .
\end{equation}
With these coordinates we find the conformal rescaling  to the flat metric 
\begin{equation}
ds^2_{\mathrm{flat}}=2dzd\bar z -2dwd\bar w =\frac{5-(x_3-y_3)^2}{2(x_3-y_3)^2 }ds^2_{S^2\times S^2}
\end{equation}
with null infinity, $\scri$, given by $x_3-y_3= 0$.
We will prefer to write the space-time in terms of real coordinates 
 written in terms of 2-component spinors as   $x^{\alpha\dot\alpha}$  with $\alpha=0,1$, $\dot\alpha=\dot 0, \dot 1$ 
on $\R^{2+2}$ by
\begin{equation}
x^{ \alpha\dot\alpha}=\frac{1}{x_3-y_3}\begin{pmatrix}
x_1-y_2&x_2+y_1\\y_1-x_2& y_2+x_1
\end{pmatrix}\, , \qquad ds^2_{\mathrm{flat}}=  dx^{\alpha\dot\alpha} dx_{\alpha\dot\alpha}\, .
\end{equation}
Here two-component spinor indices are raised and lowered with the standard skew spinor $\varepsilon_{\alpha\beta}, \varepsilon_{\dot\alpha\dot\beta}$, $\varepsilon_{01}=\varepsilon_{\dot 0\dot 1}=1$.   The flat coordinates are  clearly invariant under the $\Z_2$-action.

Thus, inside $S^2\times S^2$, we realise $\scri$ 
as %\footnote{For non-zero cosmological constant, take $\scri$ to be $y_3=0$ and then   $\scri=S^2\times S^1$.}  
\begin{equation}
\scri=\{x_3=y_3\}\simeq [-1,1]\times S^1\times S^1,
\end{equation}
with the $x_3=y_3\in[-1,1]$, and the $S^1\times S^1$ factor being the slice  $x_3=y_3$ in  $S^2\times S^2$.

%\begin{center}

%	\includegraphics[width=75mm]{scri1.pdf}
%\caption{A cartoon of finite space-time with boundary given by null infinity $\scri$}
%\end{center}

The $\Z_2$ involution  identifies  the two opposite sides of $\scri$ in $S^2\times S^2$ but is also antipodal on all three factors of $\scri$ itself.  This means that  the lightcone of a point of $\scri$ refocusses onto its antipode.  However, when thinking of $\scri$ as a boundary of flat space-time $\R^{2+2}$ we do not divide $\scri$ itself  by the $\Z_2$ involution as this would identify antipodally opposite points at $\infty$ but space-time fields might have different limits in opposite directions.

  We can see that, as noted in \cite{Mason:2005qu, Atanasov:2021oyu},
$\scri=\R\times S^1 \times S^1 $ by introducing spherical polar coordinates on each $S^2$-factor $(\theta_i,\phi_i)\in [0,\pi]\times [0,2\pi)$, with $i=1$ for the $\bx$-factor and $i=2$ for the $\by$-factor.   On $\scri$ the azimuthal coordinates  are equal, $\theta_1=\theta_2$  and $(\phi_1,\phi_2)$ parametrize the tori cross-sections.  
To relate these coordinates   to split signature analogues of Lorentzian    Bondi $\scri$ coordinates we first introduce null coordinates on the torus $(z,\tilde z)$ as analogues of the complex stereographic coordinates $(z,\bar z)$ on the sphere found in Lorentz signature.  
Now taking the $\phi_i$ modulo $2\pi$, their range can be covered by taking instead
\begin{equation}
(\phi,\tilde\phi):=\left( \phi_1-\phi_2, \phi_1+\phi_2\right) \in [0,2\pi)\times [0,4\pi) \label{double-dom}
\end{equation}
or taken modulo $2\pi$ and $4\pi$ respectively.  This asymmetric choice allows us to express  the antipodal map as  $(\theta,\phi ,\tilde \phi) \rightarrow (\pi-\theta,\phi, \tilde \phi+2\pi)$.

The coordinates $(\phi,\tilde \phi)$ are null  and we shall consider them to be coordinates on $\RP^1$ so as to expose the full Lorentz symmetry of the finite space on homogeneous  coordinates $(\lambda_\alpha,\tilde\lambda_{\dot\alpha})$ with Lorentz indices as above.  These can be related to the torus coordinates by
\begin{equation}
(\lambda_0,\lambda_1)=\left(\cos  \frac\phi2,\sin \frac\phi2\right)  \, , \qquad (\tilde\lambda_0,\tilde \lambda_1)=\left(\cos  \frac{\tilde \phi}2,\sin \frac{\tilde\phi}2\right)  \, .
\end{equation}
We introduce a Bondi-like $u$ coordinates by  first defining $u=\cot\theta_1 \in (-\infty,\infty)$.  We can then regard the full set of coordinates $(u,\lambda_\alpha,\tilde\lambda_{\dot\alpha})$
as homogeneous coordinates
being equivalent under the rescalings
%\begin{equation}
%(\mathrm{u},\lambda,\tilde \lambda)=\left(\frac u{|\lambda||\tilde \lambda|},\frac{ \lambda_1}{\lambda_0},\frac{ \tilde\lambda_{\dot 1}}{\tilde \lambda_{\dot 0}}\right)\,, \qquad |\lambda|= \sqrt{\lambda_0^2+\lambda_1^2}\, , \quad |\tilde \lambda|:=\sqrt{\tilde{\lambda}_{\dot 0}^2+\tilde \lambda_{\dot 1}^2}
%\end{equation}
%so that points of $\scri$ are invariant under
\begin{equation}
(u,\lambda_\alpha,\tilde \lambda_{\dot\alpha}) \sim (a\tilde a u,a\lambda_\alpha,\tilde a\tilde \lambda_{\dot\alpha}) \, , \qquad a\in \R^*, \quad\tilde{a}\in \R^+\, .\label{weights}
\end{equation}
That $\tilde a\in \R^+$ rather than $\R^*$ gives the double cover of $\scri$ inside $(S^2\times S^2)/\Z_2$ with the asymmetry reflecting that in \eqref{double-dom}.
This now correctly manifests the action of the split signature version  $SL(2,\R)\times SL(2,\R)$ of  the complex Lorentz spin group.  In these coordinates the action of the $\Z_2$ identification of $\scri$ in the conformal compactification $\M^{2,2}$ of $\R^{2,2}$ is  realised by $\tilde a=-1$.

If we wish to work with local coordinates we will often simply use the real null coords 
\begin{equation}
(\mathrm{u}, \lambda ,\tilde \lambda)=\left(\frac{u}{\lambda_0\tilde\lambda_0},\frac{\lambda_1}{\lambda_0},\frac{\tilde\lambda_1}{\tilde \lambda_0}\right)\, .
\end{equation}
These are good away from $\lambda_0=0$ and $\tilde\lambda_0=0$.  
We will understand an asymptotically flat split signature metric to be one that looks like our flat model to leading order near $\scri$.  This gives  
\begin{equation}
ds^2= \frac{1}{R^2}\left( d\mathrm{u} dR - d\lambda d\tilde \lambda + R\bigma d\tilde \lambda^2 + R\tilde \bigma d\lambda^2 +\ldots \right) \, , 
\end{equation}
where $R=1/r$, and $\scri=\{R=0\}$, and $\bigma(u,\lambda,\tilde\lambda), \tilde \bigma(u,\lambda,\tilde\lambda)$ are the asymptotic \emph{shears}. 
In terms of homogeneous coordinates, we rewrite this as 
\begin{equation}
ds^2= \frac{1}{R^2}\left( d u dR - D\lambda D\tilde \lambda + R\bigma D\tilde \lambda^2 + R\tilde \bigma D\lambda^2 +\ldots \right)  , \quad D\lambda:=\langle\lambda \, d\lambda\rangle\,,\;
 D\tilde \lambda=[\tilde\lambda d\tilde \lambda]\, .
\end{equation}
Counting weights under \eqref{weights} so that $u$ has weight $(1,1)$, we have that $R$ has weight $(1,1)$, $D\lambda$ weight $(2,0)$,  $\bigma$ weight $(1,-3)$ and $\tilde \bigma$ weight $(-3,1)$.

\subsection{Real and complex twistors in split signature and relations to $\scri$} 
The real twistor space $\PT_\R$ is defined to be the 3-parameter  space of \emph{$\beta$-planes}  which are totally null ASD   $S^2$s  given by\footnote{The SD $\alpha$-planes are the orthogonal transformations with negative determinant.}
\begin{equation}
\bx = A\by\, , \qquad A\in SO(3)=\RP^3:=\PT_\R\, .\label{O3-twistors}
\end{equation}
To see that $SO(3)$ is the same as real projective 3-space, $\RP^3$, note that $SO(3)=SU(2)/\Z_2$ and that $SU(2)$, unit determinant $2\times 2$ matrices is $S^3$, and $\RP^3=S^3/\Z^2$.

This relates to the usual homogeneous coordinate parametrization of $\PT_\R$ via a pair of real 2-component spinors $(\lambda_\alpha,\mu^{\dot\alpha})$ related to affine coordinates $x^{\alpha\dot\alpha}$ 
on $\R^{2+2}$ via  
\begin{equation}
\mu^{\dot\alpha}=x^{\alpha\dot\alpha}\lambda_\alpha\, , 
\end{equation}
and $\PT_\R= \RP^3$ follows by taking $(\lambda_\alpha,\mu^{\dot\alpha})$ to be  homogeneous coordinates; this   can be seen to be equivalent to \eqref{O3-twistors} after some manipulation.

%With this, $x^2=x^{\alpha\dot\alpha}x_{\alpha\dot\alpha}$

In split signature, wave functions for massless free  fields have natural representations as free data on the light-cone in momentum space, characteristic data on $\scri$ and homogeneous functions on real twistor space, $\RP^3$.  The data for the three cases are given as follows.

On momentum space we use coordinates $k_{\alpha\dot\alpha}$ which on the support of $k^2=0$ can be expressed in terms of 2-component spinors as $k_{\alpha\dot\alpha}=\kappa_\alpha\tilde \kappa_{\dot\alpha}$ and momentum space data for helicity $-n/2$ field is given by a function $\hat \phi(\kappa,\tilde \kappa)$ satisfying 
\begin{equation}
\hat \phi(a\kappa,a^{-1}\tilde \kappa)=a^{-n}\hat \phi(\kappa,\tilde \kappa)\, .
\end{equation}

 In split signature  $\scri=\R\times S^1\times S^1$ and we use homogeneous coordinates 
\begin{equation}
(u,\lambda_\alpha,\tilde \lambda_{\dot\alpha}) \sim (a\tilde a u,a\lambda_\alpha,\tilde a\tilde \lambda_{\dot\alpha}) \, , \qquad a\in \R^*\,,\quad \tilde{a}\in \R^+ ,
\end{equation}
related to standard Bondi $\scri$ coordinates by
\begin{equation}
(\mathrm{u},z,\tilde z)=(u/\lambda_0\tilde{\lambda}_{\dot 0}, \lambda_1/\lambda_0, \tilde\lambda_{\dot 1}/\tilde \lambda_{\dot 0})\, .
\end{equation}
The radiation field for a massless field of helicity $-n/2$ is a homogeneous function $\phi^0(u,\lambda,\tilde\lambda)$ of weight
\begin{equation}
\phi^0(a\tilde au,a\lambda,\tilde a \tilde \lambda)=a^{-n-1}\tilde a ^{-1}\phi^0(u,\lambda,\tilde\lambda)\, .
\end{equation} 

On twistor space $\PT_\R=\RP^3$, we use real homogeneous coordinates $Z\in\R^4$ that can be expressed as a  pair of real two component spinors $Z=(\lambda_\alpha,\mu^{\dot\alpha})$.  The incidence relation with space-time is given by
\begin{equation}
\mu^{\dot\alpha}=x^{\alpha\dot\alpha} \lambda_\alpha\, ,
\end{equation}
where  $x^{\alpha\dot\alpha}$ too is taken to be real. Helicity $-n/2$ fields then correspond to weight $-n-2$ functions
\begin{equation}
f(aZ)=a^{-n-2}f(Z)\, .
\end{equation}
We review the various twistor integral formulae in split signature and their relationship with Fourier representations and Kirchoff-d'Adhemar integral formulae in appendix \ref{int-formulae}.

\section{Twistors for split signature SD vacuum metrics and $Lw_{1+\infty}$}\label{gen-functions}

In this section we review the twistor description of \cite{Lebrun:2007} of global conformally self-dual split signature metrics.  These, if global and nontrivial, must live on $S^2\times S^2$.  We then specialize that construction to the vacuum Einstein case, introducing a Poison structure and 1-form that lead to a description in terms of generating functions.  In the last subsection we explain how $Lw_{1+\infty}$ acts both as the natural diffeomorphism symmetry of the twistor space, whereas its complexification acts transitively on the space of self-dual metrics giving it the structure of a homogeneous space.

\subsection{Conformally self-dual Zollfrei metrics}\label{Rev-confSD}
%We first recall the self-duality equations:
% on a 4-dimensional manifold $(M^4,g)$, the 2-forms split into self-%dual and anti-self-dual parts
%$
%\Omega^2_M= 
%\Omega^{2+} \oplus  \Omega^{2-}
%$ which are both real in split signature\footnote{and also in euclidean signature, but complex conjugates in Lorentz signature.}.
%With respect to this,  the Riemann curvature tensor decomposes as 
%$$ 
% \mbox{Riem}=\begin{pmatrix}
%\mbox{Weyl}^+ + S \delta & \mbox{Ricci}_0\\  \mbox{Ricci}_0 & %\mbox{Weyl}^- + S \delta
%\end{pmatrix}.
%$$
In this subsection we focus on the condition that  the Weyl tensor is self-dual.
%, $\mbox{Weyl}^-=0$.
  This implies that  $\beta$-planes survive as $\beta$-surfaces \cite{Penrose:1976jq, Penrose:1976js}, although in general the self-dual weyl curvature will obstruct the existence of $\alpha$-surfaces.   Following \cite{Lebrun:2007}, we find that such
 $\beta$-surfaces are projectively flat, and if compact, they are  necessarily $S^2$ or $\RP^2$.
In this case, null geodesics are projectively $
\RP^1$s or its double cover. Following Guillemin \cite{Guillemin:1989}, we define
\begin{defn}
A   space $(M^d,g)$, with $g$ not positive definite, is  Zollfrei if all null geodesics are embedded $S^1$s .
\end{defn}
It is proved in \cite{Lebrun:2007} that

\begin{thm}[LeBrun \& Mason]\label{Duke-thm}
Let $(M^4,[g])$ be Zollfrei with SD  Weyl tensor.  Then either
\begin{itemize}
\item   $M=S^2\times S^2/\Z_2$ with the standard conformally flat conformal structure, or
\item $M=S^2\times S^2$ and there is a  $1:1$-correspondence  between 
\begin{enumerate}

\item SD conformal structures on  $S^2\times S^2$ near flat model and
\item   Deformations $\PT_\R$ of the standard embedding  $\RP^3\subset \CP^3$ modulo reparametrizations of $\RP^3$ and $PGL(4,\C)$ on $\CP^3$. 
\end{enumerate}
\end{itemize}

\end{thm}
The deformed embedding of  $\RP^3$ is space of $\beta$ planes $\PT_\R$ and $\CP^3$ is complex twistor space.

 The data of the space-time conformal structure is therefore encoded into the location of $\PT_\R$ inside $\CP^3$. The flat $\RP^3 \subset\CP^3$ is contained in some tubular   neighbourhood $U=\RP^3\times \R^3$ where the tangent space to the $\R^3$ factor is $i\times  T\RP^3$.  In $U$, $\PT_\R$ can be expressed  as the graph of a map from $\RP^3$ to $\R^3$.  Schematically:

%\includegraphics[scale=0.7]{graph.pdf}

%\vskip-1.5in
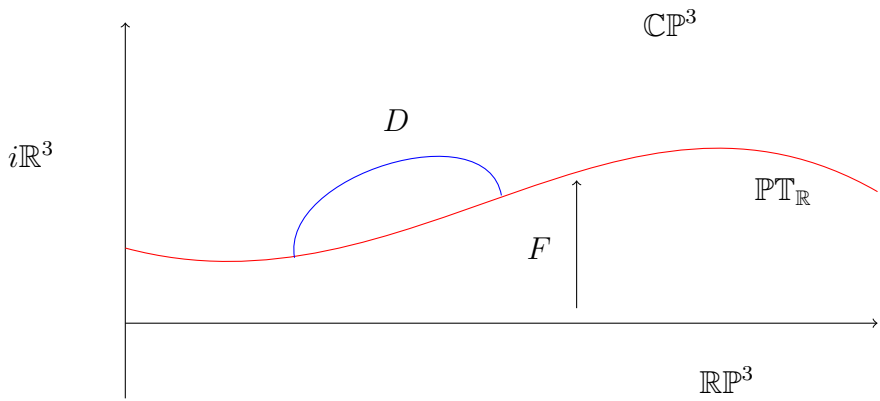
\begin{figure}[h]

\begin{tikzpicture}
	\begin{pgfonlayer}{nodelayer}
		\node [style=none] (0) at (-10, 4) {};
		\node [style=none] (1) at (-10, -1) {};
		\node [style=none] (2) at (-10, 0) {};
		\node [style=none] (3) at (0, 0) {};
		\node [style=none] (4) at (-2.75, 4) {$\mathbb{CP}^3$};
		\node [style=none] (5) at (-11.25, 2.25) {};
		\node [style=none] (6) at (-10, 1) {};
		\node [style=none] (7) at (0, 1.75) {};
		\node [style=none] (8) at (-2, -0.75) {};
		\node [style=none] (9) at (-2, -0.75) {$\mathbb{RP}^3$};
		\node [style=none] (10) at (-11.25, 2.25) {};
		\node [style=none] (11) at (-11.25, 2.25) {$i\mathbb{R}^3$};
		\node [style=none] (12) at (-11.75, 2.25) {};
		\node [style=none] (13) at (-7.75, 0.87) {};
		\node [style=none] (14) at (-5, 1.7) {};
		\node [style=none] (15) at (-6.4, 2.7) {$D$};
		\node [style=none] (17) at (-1.25, 1.75) {$\PT_\R$};
				\node [style=none] (18) at (-4.5, 1) {$F$};
	\end{pgfonlayer}
	\begin{pgfonlayer}{edgelayer}
		\draw [->] (1.center) to (0.center);
		\draw [->] (2.center) to (3.center);
		\draw [red,in=150, out=-15] (6.center) to (7.center);
		\draw [blue,in=100, out=100] (13.center) to (14.center);
	\draw [->] (-4,0.2) to (-4,1.9);
	\end{pgfonlayer}
\end{tikzpicture}
\caption{ $\PT_\R$ as graph of  $F:\RP^3\rightarrow \R^3$ inside a tubular neighbourhood $U \supset \RP^3$ in $\CP^3$.  \\ $D$ represents a holomorphic disc inside $\CP^3$ with $\p D\subset\PT_\R$.  }
\end{figure}

A condensed summary of the proof in \cite{Lebrun:2007} follows; although we dont need the first two paragaphs below in the following, we will later use  the reconstruction of space-time that arises from studying holomorphic discs in $\CP^3$ with boundary in $\PT_\R$. 

The real twistor space  $\PT_\R$ is defined to be the space of real $\beta$-surfaces in $M$.  It can be constructed as a quotient of the projective ASD spin bundle $\PS\rightarrow M$, whose fibres are $\RP^1$, the projectivisation of the real spin bundle $\bbS$. This is because  $\PS$ is foliated by the horizontal lifts of $\alpha$ surfaces in $M$ as $\PS$ can be identified with the projectivised bundle of simple ASD 2-forms inside $\Omega^{2-}$; these   determine the ASD totally null 2-plane elements tangent to the $\beta$-planes.  

To embed $\PT_\R$ inside a complex twistor space, we first    complexify this  projective  spin bundle $\CP \mathbb{S}\rightarrow M$.  This gives the $\CP^1$-bundle over $M$ obtained by complexifying the $\RP^1$ fibres of $\P$.  Then on $S^2\times S^2$, $\PS$ cuts $\CP\bbS$  into two halves $\CP \mathbb{S}^\pm$ as each $\RP^1$ divides the $\CP^1$ into two halves and  $S^2\times S^2$ is simply connected so the two halves cannot swap around\footnote{On $S^2\times S^2/\Z_2$, they do in fact swap as one traverses a generator of the $\Z_2$ and so the complement has just one component.}.  Taking $\CP\bbS^+$, we discover the that interior is already endowed with a complex structure using that of the $\CP^1$ fibres vertically, and complex ASD 2-forms horizontally.  We now blow down the boundary real slice $\PS$, taking the quotient by its foliation by $\beta$-planes.  As part of that, recall that these are necessarily projectively flat $S^2$s in the $S^2\times S^2$ case.   Each such flat projective $S^2$ has a $\Z_2$ antipodal map projecting to $\RP^2$.  Quotienting first by this ends up closing the manifold into a compact manifold without boundary.  Now blowing this $\PS/\Z_2$ down  gives a compact topological $\CP^3$ with a complex structure which must, by rigidity of the complex structure on $\CP^3$, and arguments concerning sufficient smoothness for the Newlander-Nirenberg theorem, be the standard one.  We have also on the way constructed a deformed $\PT_\R$ as  the blow down of $\PS$.  Thus we are left   with the deformation of the real slice $\PT_\R$ lying naturally inside the $\CP^3$.

For the reconstruction of $M$ from twistor space $\PT_\R$, we see from the construction that 
 each $x\in M \leftrightarrow  $  holomorphic disc   $ D_x\subset \CP^3$ with  $\p D_x \subset\PT_\R$ which is the projection of the fibre of $\CP \mathbb{S}^+$ over $x$. It can be checked that :
\begin{itemize}

\item  $D_x$ generates  the degree-1 class in $H_2(\CP^3, \PT_\R,\Z)=\Z$.

\item The moduli space of all such disks: 
$$
M= \{\mbox{Moduli of degree-1 hol. disks: } D_x\subset \CP^3, \p D_x \subset \PT_\R\} 
$$
is 4 dimensional and compact of  topology $S^2\times S^2$.

\item $M$ admits a SD conformal structure for which if the boundaries of two disks meet at a point $Z\in \PT_\R$,  $\p D_x\cap \p D_{x'}=Z$, then   $x,x'$ sit on same $\beta$-plane:

\begin{figure}[h]

\begin{center}
\includegraphics[width=90mm]{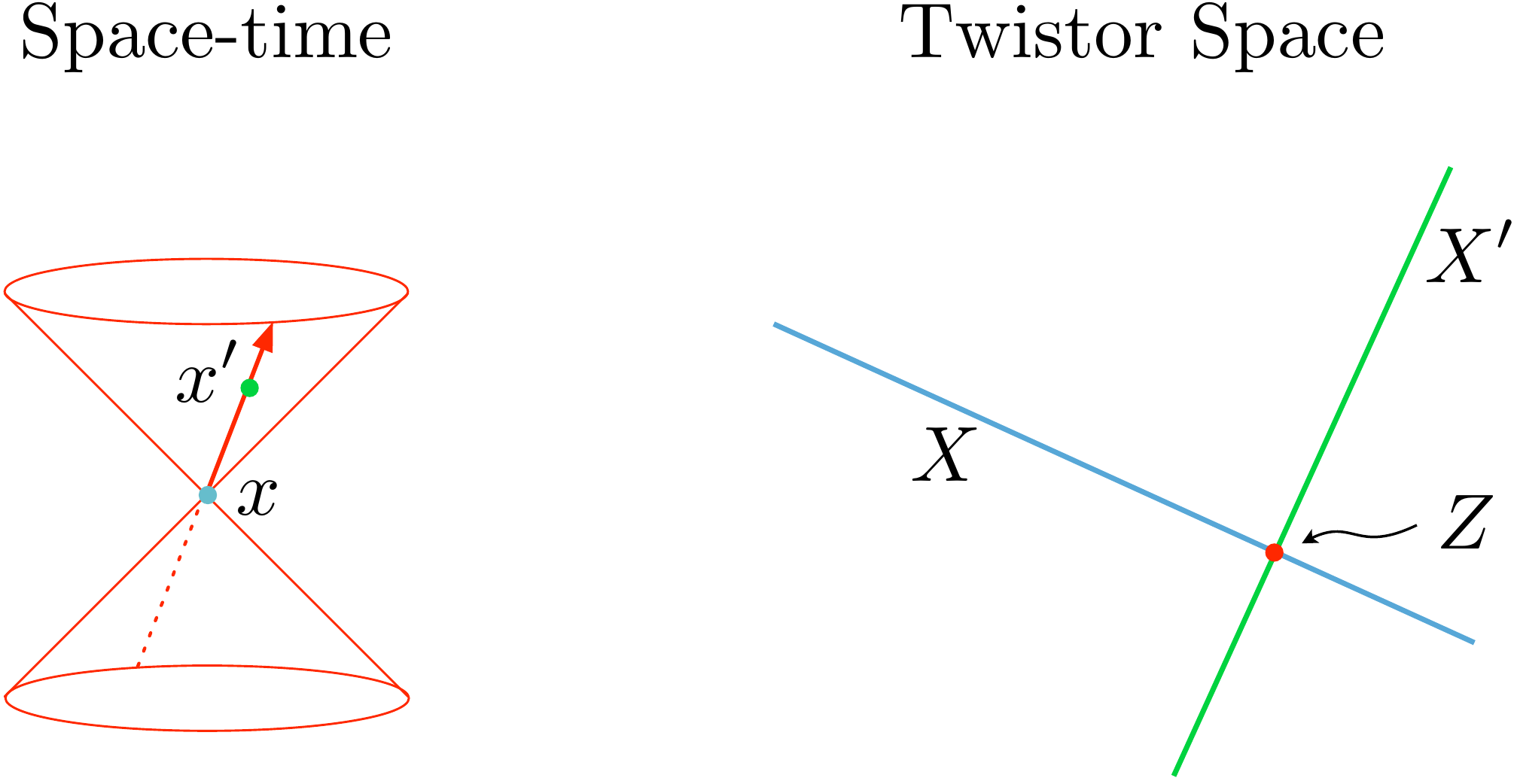}
\end{center}
\caption{Null separation between points $x,x'\in M$  corresponds to intersection between the corresponding deformed lines $X, X'$ where $X=\p D_X $ in $\PT_\R$.  This gives  a deformation of the Klein correspondence. }

\end{figure}
\end{itemize}
This completes our summary of \cite{Lebrun:2007}, to which the reader is referred for full  details.

\subsection{Generating functions for vacuum Einstein with $\Lambda =0$. }\label{gen-fns}
We now adapt the correspondence to vacuum  Einstein metrics with vanishing cosmological constant.
We will consider self-dual conformal structure  on $S^2\times S^2$ that contain a complete Einstein $g\in[g]$ with $\Lambda=0$ on the complement of the null hypersurface $\scri$,  the conformal infinity.  We will see that, at least for small data,  these can be constructed globally from  the same homogeneity degree two function $h(U)$ on $\RP^3$ as used in linear theory via the X-ray transform \eqref{X-ray-plus}, but now used in a fully nonlinear correspondence.

 To characterize the corresponding $\PT_\R\subset \CP^3$, we first break conformal invariance by introducing the \emph{infinity twistors} $I_{AB}$ and $I^{AB}$.
Let $Z^A$, $A=1,\ldots ,4$ be homogenous coordinates for $\CP^3$ that are real on $\RP^3\subset \CP^3$. We  introduce a real totally skew  $\varepsilon^{ABCD}$.  Then for vanishing cosmological constant we introduce skew twistors, \emph{infinity twistors},
\begin{equation}
I_{AB}=I_{[AB]},\quad  I^{AB}=\half\varepsilon^{ABCD}I_{CD}, \quad \mbox{ with
}\quad
I^{AB}I_{AC}=0\, .
\end{equation}

These  define a weighted 1-form  and  Poisson structure on $\CP^3$ by the equations
\begin{equation}
\theta=I_{AB}Z^AdZ^B\in \Omega^1(2)\, , \qquad \{ f, g\}:= I^{AB}\frac{\p f}{\p Z^A}\frac{\p g}{\p Z^B}
\end{equation} 
of homogeneity degree $2$ and $-2$ respectively. 
We can be more explicit by decomposing twistors into pairs of spinors in the standard way.
Let 
\begin{equation}
Z^A=(\lambda_\alpha,\mu^{\dot\alpha}), \qquad \alpha,\beta=0,1, \quad \dot \alpha,\dot \beta=\dot 0,\dot 1\, ,
\end{equation} 
and introduce $\varepsilon_{\alpha\beta}=\varepsilon_{[\alpha\beta]}
$  and $\varepsilon_{\dot \alpha\dot \beta}=\varepsilon_{[\dot \alpha\dot \beta]}
$ so that in these coordinates
\begin{equation}
\theta=\lambda_\alpha d\lambda_\beta\varepsilon^{\alpha\beta} \, , \qquad \{f ,g\}= \varepsilon^{ \dot\alpha\dot\beta}\frac{\p f}{\p \mu^{\dot\alpha}}\frac{\p g}{\p \mu^{\dot\beta}}\, .
\end{equation}
This Poisson bracket is the one that underpins $Lw_{1+\infty}$ structure as described in \cite{Adamo:2021lrv}.
It determines a  projection $p:\PT-\{\lambda_\alpha=0\}\rightarrow \CP^1$ given in homogeneous coordinates by $Z^A\rightarrow \lambda_\alpha$.  On the fibres of this projection, the Poisson structure is invertible with inverse represented by
\begin{equation}
 \omega_\lambda:=\varepsilon_{\dot\alpha\dot\beta} d\mu^{\dot\alpha}\wedge d\mu^{\dot\beta}
\end{equation}
 at fixed $\lambda$.  
The compatibility of these with the real structure guarantees that our space-time is conformal to Einstein vacuum:
\begin{thm}\label{nlg}
A vacuum  $g\in[g]$ exists iff $\lambda_\alpha$ and $\omega_\lambda$ can  both be chosen to be real on $\PT_\R$.
\end{thm}
These are sections of  $\cO(n)$ for $n=1, 2$; the real slice of these line bundles is defined via the identification  $\cO(-4)_{\R}=\Omega^3_{\PT_\R}$.  These are  analogous to Penrose's \cite{Penrose:1976js}.
We will indicate the proof later when we turn to the construction of space-time and its metric. 
 
It is easy to implement these conditions in generality with generating functions.  To set them up, decompose the complex homogeneous coordinates into real and imaginary parts, i.e., 
\begin{equation}
Z^A=U^A+iV^A\, , \qquad \mbox{ and }\qquad
\mu^{\dot\alpha}=u^{\dot\alpha}+iv^{\dot\alpha}.
\end{equation}
 Let $h(U)$ be an arbitrary function of homogeneity degree-2, 
\begin{equation}
U\cdot \frac{\p h}{\p U }=2h.
\end{equation}

\begin{propn}
 The Einstein conditions of Theorem \ref{nlg} are satisfied by   
\begin{equation}
\T_\R= \left\{ \lambda_\alpha=\bar\lambda_\alpha\, , v^{\dot\alpha}= \varepsilon^{\dot\alpha\dot\beta} \frac{\p h}{\p u^{\dot\beta}}\right\}
\end{equation}
so that projectivising gives $\PT_\R$.
 All small such deformations of $\PT_\R$ arise in this way and correspond to all small  finite vacuum  perturbations of the metric near $\scri$ that extend conformally over $S^2\times S^2$. 
 \end{propn}

 With this, it is easy to see that for $f,g$ real on $\PT_\R$, then $\{f,g\}$ is real too on $\PT_\R$. To see this, its helpful to work on the fibre $\lambda_\alpha=$ constant, on which the Poisson structure is invertible, and, whose imaginary part is 
 \begin{equation}
\Im ( \omega_\lambda)=\varepsilon_{\dot\alpha\dot\beta} d u^{\dot\alpha}\wedge d v^{\dot\beta} =\left(\frac {\p v^{\dot\gamma}}{\p u^{\dot\gamma}}\right) \varepsilon_{\dot\alpha\dot\beta} du^{\dot\alpha}\wedge d u^{\dot\beta}\, .
\end{equation}
It is now easy to see that ${\p v^{\dot\gamma}}/{\p u^{\dot\gamma}}=0$ is the integrability condition for 
$v^{\dot\alpha}= \varepsilon^{\dot\alpha\dot\beta} {\p h}/{\p u^{\dot\beta}}$.

\subsubsection*{
Construction of holomorphic discs}
As before. the split signature space-time is reconstructed as the moduli space of holomorphic discs of degree-1 with boundary on $\PT_\R$. However, these can now be constructed as sections of the fibration $p:\CP^3-\CP^1\rightarrow \CP^1$. These holomorphic discs can be parametrized in homogeneous coordinates $\lambda_\alpha$ as the sets 
\begin{equation}
\bbD=\{ \lambda_\alpha | \langle\lambda\, \bar\lambda\rangle:=\lambda_\alpha\bar\lambda^\alpha\geq 0\}/\C^*\, .
\end{equation}
where we take $\lambda_\alpha$ up to constant rescalings.  The sections of $p$ are then defined by  $(\lambda,\mu^{\dot\alpha}(\lambda)):\bbD\rightarrow \CP^3$ and this defines a holomorphic disc of degree one with boundary on $\PT_\R$ if $\mu^{\dot\alpha}$ has homogeneity degree 1 in $\lambda_\alpha$. 
The boundary condition on $ \p\bbD=\{\lambda=\bar\lambda\}$ becomes
\begin{equation}
 \mu^{\dot\alpha}(\lambda)=u^{\dot\alpha}(\lambda)+iv^{\dot\alpha}(\lambda)\, , \qquad v^{\dot\alpha}(\lambda)= \varepsilon^{\dot\alpha\dot\beta}\frac{\p h}{\p u^{\dot\beta}}(u(\lambda))\, .
\end{equation}
The general results of \cite{Lebrun:2007} show that there is a real four-dimensional moduli space $M^4$ of such discs.  We can parametrize them with coordinates $x^{\alpha\dot\alpha}$  by writing
\begin{equation}
\mu^{\dot\alpha}= x^{\alpha\dot\alpha} \lambda_\alpha +m^{\dot\alpha}
\end{equation}
where $ x^{\alpha\dot\alpha} $ are real, and $m^{\dot\alpha}$ vanishes at $\lambda_0=0$ and $\lambda_1=0$.  We can then define the map
\begin{equation}
p: M^4\times \bbD\rightarrow \PT\, ,\qquad (x^{\alpha\dot\alpha},\lambda_\alpha) \rightarrow (\lambda_\alpha,\mu^{\dot\alpha}(x,\lambda))\, .
\end{equation}
The main step in reconstructing the metric is to observe that the pullback 2-forms $p^*\omega_\lambda$, are real on $\lambda_\alpha=\bar\lambda_\alpha$ and hence extend  by reflection in $\p \bbD$  to all complex values of $\lambda_\alpha$ as 2-forms on $M^4$ of homegeneity degree 2 in $\lambda_\alpha$. Hence by an extension of Liouville's theorem we can write
\begin{equation}
p^*\omega_\lambda=\lambda_\alpha\lambda_\beta \Sigma^{\alpha\beta}
\end{equation}
where $\Sigma^{\alpha\beta}$ are a triple of real 2-forms on $M^4$ that are closed and, because $\omega_\lambda$ has rank two, satisfies 
\begin{equation}
\Sigma^{(\alpha\beta}\wedge\Sigma^{\gamma\delta)}=0\, .
\end{equation}
Contracting $\Sigma^{\alpha\beta}$ with $SO(2,1)$ Pauli matrices, $\sigma^i_{\alpha\beta}$, $i=1,2,3$, yields the three symplectic forms $\omega^i=\sigma^i_{\alpha\beta}\Sigma^{\alpha\beta}$ of a pseudo-hyperKahler structure on $M^4$
\begin{equation}
 d\omega^i=0\,, \qquad \omega^i\wedge\omega^j=\eta^{ij}\nu\,, \qquad \eta^{ij}=\diag (1,-1,-1)\, .
\end{equation}

\subsection{The  $Lw_{1+\infty}$ symmetry action on twistor space}\label{Lw1+infty}

Here we explain how the ideas of \cite{Adamo:2021lrv} adapt to this split signature context giving a real description.
In particular we will see that the symmetry algebra has a real version $Lw_{1+\infty}$, and a complex version $Lw^\C_{1+\infty}$.  The real version is the underlying structure-preserving diffeomorphism symmetry of the real  twistor space $\PT_\R$.  The complex version shifts the  location of the linear real $\RP^3$ inside $\CP^3$ taking it to $\PT_\R$.

Following \cite{Hoppe:1988gk,Bakas:1989mz} we define the classical limit $w_{\infty}$ of the $N\rightarrow\infty$ $W_N$ algebra to be the Poisson diffeomorphisms of the plane.  To make contact with our notations, let the plane have coordinates $\mu^{\dot\alpha}$, $\dot\alpha=\dot 0,\dot 1$ as before with Poisson bracket 
$$
\{ f,g\}:= \varepsilon^{\dot\alpha\dot\beta}\frac{\p f}{\p \mu^{\dot\alpha}}\frac{\p g}{\p \mu^{\dot\alpha}}\, , \qquad \varepsilon^{\dot\alpha\dot\beta}=\varepsilon^{[\dot\alpha\dot\beta]}, \quad\varepsilon^{\dot 0\dot 1}=1\, .
$$
Then a
 basis of $w_{1+\infty} $ is given by  hamiltonians that are polynomials in these coordinates, i.e.
 \begin{equation}
w^p_m=(\mu^{\dot 0})^{p-m-1}(\mu^{\dot 1})^{p+m-1}\, , \qquad |m|\leq p-1, \quad  2p-2\in\N \, .
 \end{equation}
The
 Poisson brackets then give rise to the  commutation relations  of $w_{1+\infty}$:
$$
\{w^p_{m},w^q_{n}\}=(2(p-1)n-2(q-1)m)w^{p+q-2}_{m+n}\, ,
$$
with the $1+$ denoting the central constant Hamiltonian.

To obtain the full structure preserving diffeomorphisms of twistor space, we need the loop algebra $Lw_{1+\infty}$, where the loop coordinate is   $\frac{\lambda_1}{\lambda_0}=\tan \frac{\theta}2$.  This loop algebra has  generators \begin{equation}
g^p_{m,r}=w^p_m \e^{ir\theta}\,, \qquad r\in \Z .
\end{equation}
With Poisson brackets now
\begin{equation}
\{g^p_{m,r},g^q_{n,s}\}=\left(2(p-1)n-2(q-1)m\right)g^{p+q-2}_{m+n,r+s}\, ,
\end{equation}
giving the standard commutation relations of $Lw_{1+\infty}$.

It is now straightforward to see these are the structure preserving diffeomorphism group of $\PT_\R$ preserving the structures of the split signature nonlinear graviton theorem \ref{nlg}.  Furthermore,  the complexification $Lw^\C_{1+\infty}$ of $Lw_{1+\infty}$ can be understood to act as the Lie algebra of a semigroup generated by holomorphic diffeomorphisms defined near $\RP^3$ inside $\CP^3$.  These are generated by complex hamiltonians and serve to shift the location of $\PT_\R$ inside $\CP^3$ thus generating nontrivial gravitational data (the real diffeomophism are of course just  coordinate transformations).  Thus the gravitational data can be defined to be these complex transformations defined modulo the real ones realizing the space of self-dual gravitational data as $Lw^\C_{1+\infty}/Lw_{1+\infty}$.  Our generating functions $h(U)$ provide natural representatives of these cosets via a choice of pure imaginary slice.

\section{From twistor data to  asymptotic shears on $\scri$  and back}\label{transform-scri}
%Near $\scri$, our metrics look like
%$$
%ds^2=\frac{1}{(x_3-y_3)^2}( ds^2_{S^2_\bx} - %ds^2_{S^2_\by} )+ \ldots\, , 
%$$ 
%in coordinates  $(\bx,\by)\in \R^3\times \R^3$, $|\bx|=|\by|=1$.  These coordinates relate to   analytically continued Bondi coordinates $(u,\lambda_\alpha,\tilde \lambda_{\dot\alpha})$ by
%\begin{equation}

%\end{equation}  

We briefly sketch the  construction of the generating function $h(U)$ from the gravitational data, the asymptotic shear $\bigma$, at $\scri$, firstly for  the fully nonlinear regime    and then for the linearized transform. There is a nontrivial aspect to the latter as the linear correspondence \eqref{scri-radon} gives  shears $\bigma$ that are even under the $\Z_2$ but we address this second as it provides further insight into the nonlinear discussion.

\subsection{The nonlinear case}
 To construct $h(U)$ from the $\scri$-data, in the nonlinear regime,we can apply  a lower dimensional analogue of Theorem \ref{Duke-thm} given by Lebrun 
and the author in \cite{Lebrun:2002}. 
 In this theorem, Zoll
 %\foonote{i.e., all of whose geodesics are embedded $S^1$s.}  
 projective structures are considered on $\RP^2$ and $S^2$.  A projective structure is an equivalence class of torsion-free connections with two connections equivalent if they have the same geodesics.  The Zoll condition, like the Zollfrei condition is that  all of their geodesics are embedded circles.  Those on $\RP^2$ are rigid, whereas for $S^2$ we have
\begin{thm}[Lebrun \& Mason] \label{proj-surfaces}
There is a $1:1$ correspondence between Zoll projective structures on $S^2$ close to the standard one, and deformations $\PT_\R^\lambda$ of the embedding of the standard real slice $\RP^2$ of $\CP^2$. 
\end{thm}  

In our context,  the real twistors in our $(2,2)$ space-time that correspond to points of $\PT_\R$ intersect $\scri$ in null geodesics in the planes $\lambda_\alpha=$ constant.  We can apply the above theorem for each such fixed real $\lambda$ to give a deformed real slice $\PT_\R^\lambda
$ inside each $\PT^\lambda$, the $\CP^2$ inside $\PT$ whose tangent plane is determined by the covariantly constant spinor $\lambda$.  Recalling that a self-dual Ricci-flat space-time admits covariantly constant spinors, such planes in $\PT$ will be defined globally in the complex in our context and by globality take the standard flat space form.

  In the 
flat case, the null geodesics are lines given by $u=\mu^{\dot\alpha}\tilde \lambda_{\dot\alpha}$.  however, for $\bigma\neq 0$ the null geodesic equation on $\scri$ is nontrivial, defining  a projective structure. At fixed $\lambda_\alpha$, the null geodesics are curves $u=Z(\tilde \lambda)$ in $\scri$ satisfying\footnote{Although this equation is written in terms of affine coordinates, in homogeneous coordinates, $Z$ has weight 1 in $\tilde \lambda_{\dot\alpha}$ and so $\lambda_{\dot\alpha}\p^2 Z/\p\tilde\lambda_{\dot\alpha}\p\tilde\lambda_{\dot\beta}=0$ by homogeneity.  Thus  $\p^2 Z/\p\tilde\lambda_{\dot\alpha}\p\tilde\lambda_{\dot\beta}=\tilde \lambda^{\dot\alpha} \tilde\lambda^{\dot\beta} \tilde \eth^2 Z$ defines a Mobius invariant operator $\tilde \eth^2$ as in \cite{Sparling:1990,Eastwood:1981jy}. This agrees with $\p_{\tilde \lambda}^2$ for choice of affine coordinate $\tilde \lambda$. } \cite{Hansen:1978jz}
\begin{equation}
\p^2_{\tilde \lambda} Z=\bigma(Z,\tilde\lambda,\lambda).\label{good-cut}
\end{equation}
This is Newman's good cut equation \cite{Newman:1976gc} analytically continued to become a real equation in split signature.    
Globally, if the finite space-time is assumed to be simply connected, the $\alpha$-planes will necessarily  be topologically discs, and will therefore intersect $\scri$ in circles.   
 Thus Theorem \ref{proj-surfaces} can be applied at each value of $\lambda$.   This gives a deformed $\PT_\R$ that lies inside the hypersurface $\lambda_\alpha=\bar\lambda_\alpha$.

 To see that the hypersurface satisfies the second condition of Theorem \ref{nlg}, we note that the projective structure defined by \eqref{good-cut} is special in that it admits a constant Wronskian.  For a solution $Z$ to \eqref{good-cut}, perturbations $z_1, z_2$ satisfy
\begin{equation}
\p_{\tilde\lambda}^2 z_i=\dot \bigma(Z,\lambda,\tilde\lambda) z_i\, . \label{good-cut-pert}
\end{equation} 
These 
 represent two tangent vectors to $Z$ now thought of as  the corresponding point of $\PT_\R$.  Then we define $\omega_\lambda(z_1,z_2)$ by the Wronskian $z_1\p_{\tilde\lambda} z_2-z_2\p_{\tilde\lambda}$,  which becomes in these homogeneous coordinates
\begin{equation}
\omega_\lambda(z_1,z_2):=\varepsilon_{\dot\alpha\dot\beta} \frac{\p z_1}{\p\tilde \lambda_{\dot\alpha}}\frac{\p z_2}{\p \tilde\lambda_{\dot\beta}}\, .\label{omega-wronsk}
\end{equation} 
It is standard that such a Wronskian should be constant along the the geodesic and so defines a real two-form on $\PT_\R$ at fixed $\lambda$.

We must now show that $\omega_\lambda$ extends holomorphically over $\PT^\lambda$.  This will be enough to determine the generating function.  In order to do so, we first introduce some of the ingredients of \cite{Lebrun:2002}.

It follows from the discussion in \cite{Lebrun:2002} that,   the complement of the real slice in $\PT^\lambda$ should be identified  with complement of the real slice of one half $\P T_\C^+\scri_\lambda$ of the projective complexified tangent bundle $\P T_\C\scri_\lambda$ of $\scri_\lambda$; the fibres of this are $\CP^1$s that are cut in half by their real slice.     We can introduce   two real coordinates $(\tilde \lambda, u)$ and the complex fibre coordinate $v$ representing the complex vector field $\p_{\tilde \lambda} + v\p_u$ so that the geodesic flow associated to \eqref{good-cut} is
\begin{equation}
\mathcal{D}= \p_{\tilde\lambda}+v \p_u + \bigma\p_v\, .
\end{equation}
We can express this on the nonprojective tangent bundle of $\scri_\lambda$ by introducing the homogeneous coordinates $\mu^{\dot\alpha}=\p Z/\p\tilde \lambda_{\dot\alpha}$ as above, then  $(\mu^{\dot\alpha}, \tilde \lambda_{\dot\beta})$ define homogeneous coordinates on the projective tangent bundle of 2-plane $\scri_\lambda$ at fixed $\lambda$.  This follows because the homogeneity relation $Z=\tilde \lambda_{\dot\alpha}\p Z/\p\tilde\lambda_{\dot\alpha}=Z$ determines $Z$, and then the other component of $\mu^{\dot\alpha}$, its first derivative with respect to $\tilde \lambda$; in affine coordinates this gives the coordinate $v$.  The geodesic flow above together with the Euler vector field in $ \tilde \lambda_{\dot \alpha}$ can now be written in terms of these homogeneous coordinates as
\begin{equation}
V^{\dot\alpha}=\frac{\p}{\p\tilde\lambda_{\dot\alpha} }-\bigma\tilde \lambda^{\dot\alpha}\tilde \lambda^{\dot\beta} \frac{\p}{\p\mu^{\dot\beta}}\, .
\end{equation}

The real twistor space at fixed $\lambda$, $\PT^\lambda_\R$ is obtained by blowing down the real projective tangent bundle  of $\scri$, with $v=\bar v$,  along the geodesic flow $\cD$.  This is lifted to the non projective space by including also  the Euler vector field for $\tilde \lambda_{\dot\alpha}$ which combindes to give $V^{\dot\alpha}$.

Complexifying the fibre $v$ of the bundle projective  tangent bundle, we can define  the distribution
$T^{0,1}=\{\mathcal{D},\p/\p\bar v\}$.  When $v \neq \bar v$, this defines a complex structure.\footnote{ When $v=\bar v$, $\cD=\bar \cD$ so $T^{0,1}$ has a nontrivial real subbundle spanned by the geodesic flow $\cD$.}  We must show that \eqref{omega-wronsk}  extends holomorphically over the complex twistor space at fixed $\lambda$  in the context of theorem \ref{proj-surfaces}.  Our coordinates $\mu^{\dot\alpha}$ 
above naturally complexify to  the projective tangent bundle of $\scri_\lambda$ with complexified fibres; it is this space away from the real slice that  becomes  identified with the $\PT^\lambda-\PT^\lambda_\R$,  and when the real slice is blown down, this will in turn become the  plane, $\PT^\lambda=\CP^2$, at a fixed real $\lambda$. On this space the complex structure is determined by the $(0,1)$ vectors $\p/\p\bar\mu^{\dot\alpha}$ and $V^{\dot\alpha}$.  The same computation that shows that the Wronskian is constant now shows on this partially complexified  space that \eqref{omega-wronsk} is holomorphic and so extends holomorphically over the whole $\CP^2_\lambda$ as  required.
This is sufficient to determine $h(U)$ on each each $\CP^2_\lambda$ for each real $\lambda$ via the argument used above on $\CP^3$ and hence we have reconstructed $h(U)$.

\subsection{Linear case}
The issue here is that the generalised Radon or Funk  transform \eqref{scri-radon} leads to shears that satisfy the opposite parity in $\tilde \lambda$  to that appropriate to its weight
\begin{equation}
\bigma(u,\lambda,\tilde \lambda)= \bigma(-u,\lambda_\alpha,-\tilde \lambda_{\dot \alpha})\, .\label{even-shear}
\end{equation}
This follows from the fact that a sign flip for $\tilde \lambda_{\dot\alpha}$ flips the sign of $u$ as $u=\mu^{\dot\alpha}\tilde\lambda_{\dot \alpha}$ but can be absorbed into a sign flip in the parameter $t$ in the integral as the sign flip in $dt$ is compensated by the flip in orientation of the $t$ line.\footnote{ Since $\bigma$ has weight $-3$ in $\tilde \lambda_{\dot\alpha}$, this is the opposite behaviour to what one might expect based on weight, but is turns out to be sufficient (and necessary) to guarantee that to first order the solutions to \eqref{good-cut} are circles. To see this we can note that we can rewrite the first order perturbation of  \eqref{good-cut} as an indefinite integral equation for the perturbation $z =\delta Z(\tilde\lambda)$ of the flat $Z=[\mu \tilde \lambda]$ by
\begin{equation}
z(\tilde{\lambda})=\int^{\tilde\lambda}_{\tilde\lambda_0} [\tilde \lambda \tilde \lambda'] \bigma([\mu\tilde\lambda'],\lambda,\tilde\lambda') [\tilde \lambda' d\tilde\lambda']\, .\label{odd-sigma}
\end{equation}
this integrand is odd under the antipodal map and so integrates to zero giving solutions $z$ that are single-valued.  }
One might therefore be concerned that the characteristic data is therefore special.  
However,  at least for the case of the ultrahyperbolic wave equation on $\R^4$, the X-ray transform range is provided to be surjective under reasonable conditions in \cite{Gelfand:2003}.  The equation \eqref{odd-sigma} is precisely a consequence on the restriction of the X-ray transform to $\scri$ where it degenerates into the Radon transform \eqref{Radon}. Similarly the half-Fourier transform between momentum space and twistor space also gives surjectivity and that leads to \eqref{Radon} and hence \eqref{odd-sigma}.  Geometrically, this condition is necessary and sufficient to guarantee that the null geodesics at $\scri$ formed by intersections with $\beta$-planes are circles to first order around the projectively flat case  \cite{Lebrun:2002}.

\section{From open sigma models to scattering  amplitudes}

Following the twistor strings of Witten \cite{Witten:2003nn}, Berkovits \cite{Berkovits:2004jj} and Skinner \cite{Skinner:2013xp}, curves of higher degree can be expected to give rise to   scattering amplitudes of higher MHV degrees. The reconstruction of space-time as a moduli space of holomorphic discs with boundaries on $\PT_\R$ of degree-one therefore motivates the study of more general holomorphic disks of higher degree for the construction of amplitudes.  Although we could consider the fully quantized twistor-string with target supertwistor space, we instead  adapt the chiral sigma model introduced in \cite{Adamo:2021bej}  to  introduce an open version with boundary conditions provided by the real slice $\PT_\R$.  Unlike the twistor-string (or ambitwistor-string), this is non-supersymmetric, is  treated classically  but does not involve  dual twistors, living entirely in twistor space.  It is more directly related to the above real geometry and allows for a simpler formulation of amplitudes on a fully nonlinear self-dual background as in \cite{Adamo:2022mev}. 

\subsection{The open twistor sigma model}\label{Open-model}

%We here introduce an action for the holomorphic disks that on-shell will provide a functional that generates amplitudes as a split signature analogue of \cite{Adamo:2021bej}.  
In order to allow higher MHV-degree, we will need to insert $k$ ASD twistor functions $\tilde{h}_i$ of homogeneity degree $-6$ defined in the first instance on\footnote{When expanded around flat space, $\PT_\R$ will be the standard $\RP^3\subset\CP^3$. } $\PT_\R$. These define the ASD perturbations and are      inserted  at $k$ points $Z_i\in \PT_\R$, $i=1,\ldots ,k$.
Our
holomorphic disks $Z(\sigma): \bbD \rightarrow \T$ must not only have their   boundaries on $\PT_\R$, but must also pass through the points $Z_i$.  We can  impose this latter condition by setting
\begin{equation}
Z(\sigma)=\sum_{i=1}^k \frac{Z_i}{\sigma- \sigma_i} +M(\sigma)\label{marked-points}
\end{equation}
   where  $M(\sigma)$ is smooth.  To make $Z(\sigma)$ regular, one can consider  $Z(\sigma)\prod_i (\sigma - \sigma_i)$ which is smooth, of degree $k-1$ and passes through the $Z_i$.  It is also manifestly real when $\sigma $ is real if $M(\sigma)=0$  on the boundary thus satisfying the boundary conditions when $h=0$.  For $h\neq 0$ we must have when $\sigma=\bar\sigma$ 
\begin{equation}
%\dbar \mu^{\dot\alpha}=0\, , \qquad 
 M(\sigma)=(0,m^{\dot\alpha})\, , \qquad \Im \, m^{\dot\alpha}|_{\sigma=\bar\sigma}=\left.\varepsilon^{\dot\alpha\dot\beta}\frac{\p h(U(\sigma))}{\p u^{\dot\beta}}\right|_{\sigma=\bar\sigma}\, .\label{boundary}
\end{equation}
This will have  a unique solution  $M(\sigma)=(0,m^{\dot\alpha}(\sigma))$ holomorphic and smooth on the boundary as holomorphic functions that are holomorphic and bounded on the upper half plane and real on the real axis are constant, but must vanish at infinity if it it not to have a pole there. This boundary value problem arises from  action:
\begin{equation}
S_\D[Z(\sigma),Z_i,\sigma_i]:= \Im \int_\D [ \mu \,\dbar \mu ] d\sigma + \oint_{\p \D} h(\Re Z) d\sigma + \frac1{2\pi }\sum_i [\mu(\sigma_i) \mu_i]
\end{equation}
It can now be checked that the equations of motion give \eqref{marked-points} with $M(\sigma)$ holomorphic, together with the boundary condition \eqref{boundary}.

We will now express scattering amplitudes in terms of the \emph{on-shell action}
\begin{equation}
S^{os}[h, Z_i,\sigma_i]:= S_\D[Z(\sigma),Z_i]
\end{equation}
where the right-hand side is evaluated on the unique solution to \eqref{marked-points} with $M(\sigma)$ holomorphic satisfying the boundary condition \eqref{boundary} and so the left hand side is thought of as a functional of $h$ together with the boundary data $Z_i,\sigma_i$.
In this form we can construct amplitudes for the scattering of $k$ negative helicity (anti-self-dual) gravitons with twistor wave-functions  
\begin{equation}
\tilde h_i\in C^\infty (\PT_\R,\cO(-6)), \qquad  i=1,\ldots, k
\end{equation}
 on the self-dual background determined by nonlinear self dual data $h\in C^\infty (\PT_\R, \cO(2))$ as described above.  An amplitude in the first instance will be a functional $\mathcal{M}[h, \tilde h_i]$ of this gravitational data. The main claim is

\begin{propn}[Adapted to split signature from \cite{Adamo:2021bej}]\label{background-field-gamps}
The amplitude for $k$ ASD perturbations on SD background $h$ is
\begin{equation}
\CM(h, \tilde h_i)= \int_{(S^1\times \PT_\R)^k} S^{os}[h, Z_i, \sigma_i] \det{}' \widetilde \HH \prod_{i=1}^k \tilde h_i(Z_i) D^3 Z_i d\sigma_i\, .\label{amp-background}
\end{equation}
Here $S^{os}[h, Z_i, \sigma_i]$ is the  sigma model action evaluated on-shell, and $\widetilde \HH$ is  the $k\times k$  matrix
\begin{equation}
\widetilde{\mathbb{H}}_{ij}(Z_i)=\begin{cases} \quad\frac{\langle \lambda_i \lambda_j\rangle}{\sigma_i-\sigma_j}\qquad\qquad i\neq j\\
-\sum_l \frac{\langle\lambda_i \lambda_l\rangle}{\sigma_i-\sigma_l}\, , \qquad i=j\, .
\end{cases} \label{tilde-Hodge}
\end{equation}
and
  $\det{}' \widetilde\HH$ is the reduced determinant, i.e., that of any $k-1\times k-1$ minor with appropriate sign and $D^3Z_i:= \varepsilon_{ABCD}Z_i^AdZ_i^B\wedge \ldots \wedge dZ_i^D$.
  %defined below, of
%and the $\det{}'\tilde \H$ is the reduced determinant defined below.
\end{propn}

We first explain why \eqref{amp-background} is well-defined.  Firstly, the Mobius invariance can be understood by expressing the various quantities as sections of the $\cO(n)$ line bundles over $\D$ whose sections can be represented by  functions homogeneous degree $n$ in the homogeneous coordinates $(\sigma_0,\sigma_1)$ on $\D$.  Then in \eqref{marked-points} replace $\sigma-\sigma_i$ by $(\sigma\,\sigma_i):=\sigma_0\sigma_{i1}-\sigma_1\sigma_{i0}$ and we see that $Z(\sigma)$ as defined is a section of $\T\otimes \cO(-1)$ and $Z_i$ a section of $\T\otimes \cO_i(1)$ where $\cO_i(n)$ is the line bundle of functions of homogeneity degree $n$ on the $\sigma_i$ disc.
%With this the weights of the $\widetilde\HH_{ij}$ can be seen to be weightless in the $\sigma_i$ with vanishing row and column sums, so that the reduced determinant is well-defined.  
The weights $-6$ of the $\tilde h_i$ are then balanced by the $+4$ of the $D^3Z_i$ and the $+2$ of the $d\sigma_i$ which become $(\sigma_i d\sigma_i)$ in the homogeneous coordinates.  We also see that the entries of $\HH$ are weightless and by construction it has kernel $(1,1,\ldots ,1)$ so that the reduced determinant is well-defined.

\noindent
{\bf Proof:} 
We will be brief as the argument is essentially an adaptation of that in \cite{Adamo:2021bej} to our split signature context.   The argument follows by expanding  $h=h_{k+1}+\ldots +h_n$ to 1st order in each momentum eigenstate $h_i$  to give a standard well-known formula for the
flat background perturbative gravity amplitude \cite{ Cachazo:2012kg,Cachazo:2012pz}.

For eigenstates of momentum  $k_{i\alpha\dot\alpha}=\kappa_{i\alpha}\tilde\kappa_{i\dot\alpha}$  we have the formulae \cite{Roiban:2004yf, Witten:2004cp}
\begin{equation}
h_i=\int_R \frac{dt_i}{t_i^3}\delta^2(t_i \lambda_\alpha-\kappa_{i\alpha}) \e^{it_i[\mu,\tilde \kappa_i]}, \quad \tilde h_i=\int_\R \frac{dt_i}{t_i^{-5}}\delta^2(t \lambda_\alpha-\kappa_{i\alpha}) \e^{it_i[\mu,\tilde \kappa_i]}\, .\label{mom-estate}
\end{equation}
We can perturbatively expand the on-shell action  by setting $h=h_{k+1}+\ldots +h_n$ and expanding to 1st order in each momentum eigenstate $h_i$.  This can be expressed as a tree correlator
$$
S^{os}_D[ h_{k+1}+ \ldots + h_n, Z_i, \sigma_i]= \langle V_{h_{k+1}} \ldots V_{h_n}\rangle_{tree}  + O(h_i^2)\, ,
$$
where the `vertex operators' are
$
V_{h_i}= \int_{\p D} h_i (\sigma_i) d\sigma_i\, .
$
This a sum over connected tree diagrams on the $n-k$ vertices provided by each vertex operator with 
 propagators arising from the kinetic term in $S_\D$.  This gives the Poisson bracket $\{\, , \}$ leading, on momentum eigenstates, to 
\begin{equation}
\langle h_i h_j\rangle_{tree}=\frac{\{   h_i\,  h_j\}}{\sigma_i-\sigma_j}=\frac{t_it_j[i\, j]}{\sigma_i- \sigma_j}h_ih_j\, , \qquad i\neq j\, .
\end{equation}
Summing over all such tree-diagrams, the matrix-tree theorem then gives  the result as a reduced determinant
\begin{equation}
\langle {h_{k+1}}\ldots {h_n}\rangle_{\mathrm{tree}} =\det{}'\mathbb{H}\prod_{i=k+1}^n h_i \, , 
\end{equation}
where the Laplace matrix associated to the tree graphs is  
\begin{equation}
 \HH_{ij}=\begin{cases}\frac{t_it_j[ij]}{\sigma_i-\sigma_j}\, , \qquad\qquad i\neq j \\
 -\sum_l \frac{t_it_l[il]}{\sigma_i-\sigma_l} \, , \quad i=j
 \end{cases}
\end{equation}
This leads to
\begin{equation}
 \CM(h_i,\tilde h_i)= \int_{(S^1)^n\times (\RP^3)^k} \hspace{-1cm} \det{}' \HH  \det{}' \tilde \HH \prod_{j=k+1}^n h_j d\sigma_j \prod_{i=1}^k \tilde h_i(Z_i) D^3 Z_i d\sigma_i \, .
\end{equation}
This is now equivalent to the Cachazo-Skinner formula  \cite{ Cachazo:2012kg,Cachazo:2012pz} adapted to split signature. $\Box$

\smallskip
This sum over tree graphs was first observed in gravity amplitudes at MHV in \cite{ Bern:1998sv,Nguyen:2009jk} and the reduced determinants in \cite{Hodges:2012ym} and the comparison using the matrix-tree theorem in \cite{Feng:2012sy}.   For a more full discussion of the matrix-tree theorem in this context see \cite{Adamo:2021bej, Adamo:2022mev}.

\subsection{Invariance under  $Lw_{1+\infty}$ and vertex operators and Ward identities}
The framework, and in particular the amplitudes formulae, gives full expression of the $Lw_{1+\infty}$ symmetries.
We saw above that in split signature
 $Lw_{1+\infty}$ acts on $\PT_\R$ in two ways:
\begin{enumerate}
\item $Lw_{1+\infty}$ acts passively as the structure preserving diffeomorphisms (or coordinate transformations) of $\PT_\R$, 
\item when multiplied by $i$, $iLw_{1+\infty}$ shifts the embedding of  $\PT_\R$ inside $\CP^3$.  

\end{enumerate} 

  These sigma models encodes  the Poisson structure that underlies the $Lw^\C_{1+\infty}$ symmetry in their tree-level OPE
\begin{equation}
g(Z(\sigma))\cdot g'(Z(\sigma'))\simeq \frac{\{g,g'\}(Z(\sigma))}{\sigma-\sigma'}+\ldots\, ,
\end{equation}
for $g,g'$ local holomorphic functions on twistor space.\footnote{Quantum mechanically of course it would become arbitrarily singular.}
As far as the bulk sigma model is concerned, the local   $Lw_{1+\infty}^\C$ symmetries of the action are holomorphic Poisson vector fields with Hamiltonians given by  holomorphic functions $g(Z)$ of homogeneity degree 2.   Noether's theorem for our real version of the sigma model gives  the charge
\begin{equation}
Q_g=\Im \oint gD\sigma\, .
\end{equation}
As can be seen, if the contour  is taken to the boundary of the disc, it vanishes when $g$ is real there, as expected for a passive transformation, but it becomes  nontrivial giving the  graviton vertex operator $V_h$ where $h=\Im g$ when $g$ is imaginary.

% On the other hand, the symmetries of type one, as  the structure preserving diffeomorphisms of $\PT_\R$., i.e., those diffeomrphism preserving the fibration to $\lambda_\alpha$ and the Poisson structure.   These are generated by the Hamiltonian vector fields of hamiltonians $g(\lambda_\alpha,\mu^{\dot\alpha})$ given as smooth real sections of $\cO(2)$.  

The correspondence with $\scri$ allows us to recognize the transform to twistor space of certain BMS symmetries. 
In particular, the supertranslations $u \rightarrow u+Z(\lambda,\tilde \lambda) $ that are linear in $\tilde \lambda_{\dot\alpha}$, $ Z=Z^{\dot\alpha}(\lambda)\tilde \lambda_{\dot\alpha}$ can be realized with the Hamiltonian $g=  \mu^{\dot\alpha}  Z_{\dot\alpha}(\lambda)$.  Similarly for the super-rotations 
\begin{equation}
(\lambda_\alpha,\tilde\lambda_{\dot\alpha})\rightarrow (\lambda_\alpha, L(\lambda)_{\dot\alpha}^{\dot\beta}\tilde \lambda_{\dot\beta}), \qquad \leftrightarrow \qquad g= L_{\dot\alpha\dot\beta}(\lambda) \mu^{\dot\alpha}\mu^{\dot\beta}\, . 
\end{equation}
On the other hand, as noted above, in this signature, imaginary or complex such hamiltonians correspond to perturbations of $\PT_\R$ that shift its location inside $\CP^3$, and hence perturbations of the SD metric, i.e.\ gravitons, rather than passive coordinate transformations. These formulae underpin the connection between leading and subleading soft theorems and Ward identities for symmetries as described in \cite{Strominger:2017zoo} as they are the two terms that arise when the momentum eigenstates \eqref{mom-estate} are expanded in small momenta, see \cite{Adamo:2021lrv} for a more extended discussion. 

 Higher order soft limits were shown to agree with those of the \cite{Strominger:2021lvk} in \cite{Adamo:2021lrv} and a completely analogous set of computations could be performed here.  The main point of correspondence to note is that the $g$ or $h$ used here should be understood as the \v Cech versions of the twistor data in \cite{Adamo:2021lrv} and the $h$ used here is the result of the the light-ray transform discussed by Sharma in \cite{Sharma:2021gcz} interpreted as part of the half-Fourier transform of \eqref{half-fourier}.

%The soft expansion leads to a  mode expansion for $w_{1+\infty}$
%$$
%g_{\omega,\kappa,\tilde \kappa}=\frac{\langle\lambda o\rangle^3}{\langle \lambda\kappa\rangle}\e^{ \frac{i \omega_i [\mu \tilde \kappa]}{\langle\lambda o\rangle}} \, ,\qquad \gamma_{\kappa_i} = \{ |\lambda \lambda \kappa_i\rangle/\langle\lambda o\rangle|=\epsilon\}
%%$$

%On $\{\langle \lambda o\rangle\neq 0 \}$: $\; \dbar  g_{\omega,\kappa,\tilde \kappa}=h_{\kappa} \quad \mbox{ so }\quad Q_{g_\kappa}=V_{h_\kappa}$.
%\smallskip

%(cf.\ Sharma's light-ray transform  \cite{Sharma:2021gcz}).

 % Mellin transform to $\Delta=k \in \Z_{\leq 2}\leftrightarrow$ soft expansion in $\omega \leadsto$ 
%$$
%g^k= \frac{\mathrm{i}^{2-k}}{(2-k)!}\,\frac{[\mu\,\tilde \kappa]^{2-k}\langle\lambda o\rangle ^{k+1}}{\langle\lambda\,\kappa\rangle}\,.
%$$

%Expands to give Strominger's $Lw_{1+\infty}$-algebra of symmetries
% complete with combinatoric factors:
% $Lw_{1+\infty} \leftrightarrow$ 
%  Hamiltonians on the $\mu^{\dot\alpha}$-plane, Laurent in $\lambda$:
%$$
%w^p_{m,r}=(\mu^{\dot 0})^{p-m-1}(\mu^{\dot 1})^{p+m-1}\frac{\langle \lambda\iota\rangle^r}{\langle \lambda o\rangle^{r+2p-4}}\, , \qquad |m|\leq p-1 \qquad  
%$$
%Poisson brackets
%$$
%\{w^p_{m,r},w^q_{n,s}\}=(2(p-1)n-2(q-1)m)w^{p+q-2}_{m+n,r+s}\, .
%$$
%Generalizes supertranslations at p=2.

\section{Discussion}\label{discussion}
We have seen, building on \cite{Lebrun:2007}, that are    SD split signature conformal classes of metrics on $S^2\times S^2$ conformal to  SD  vacuum metrics on the complement of $\scri=S^1\times S^1\times \R \subset S^2\times S^2$ cutting the space into two.  They depend on the choice of an arbitrary  free homogeneity degree 2 twistor function $h(U)$ on $\RP^3$.  In the linear limit they are in 1:1 correspondence with  all asymptotically flat split signature vacuum metrics on $\R^4$ by the surjectivity of the X-ray transform \cite{Gelfand:2003}.
Geometrically, $h(U)$   defines a deformed real slice $\PT_\R$ inside $\CP^3$  via the natural Poisson structure on $\CP^3$;
space-time is reconstructed by considering open holomorphic discs of degree-1 inside $\CP^3$ whose boundary lies in $\PT_\R$.  These can be described as solutions to a chiral open sigma model with boundary on $\PT_\R$.  At higher degree, the on-shell action can then be used  to  compute gravity amplitudes at any corresponding $MHV$-degree. 

Similar results follow for $\Lambda\neq 0$ where now the double covered $\scri$ has topology $S^2\times S^1$ with a smooth Lorentzian conformal structure.  The twistor correspondence extends so that twistor functions  $h(Z)$ now determine deformations of the   conformal structure of $\scri=S^2\times S^1$ that are Zollfrei in the original sense of Guillemin \cite{Guillemin:1989}, i.e., the null geodesics are all circles. In this case the infinity twistor $I_{AB}$ is non-degenerate of rank four and the contact structure $\theta$ is therefore also non-degenerate $\theta\wedge d\theta \neq 0$.  The analogue of the $Lw_{1+\infty}$ is now deformed to become the contact structure-preserving diffeomorphisms for this non-degenerate contact structure, again determined by Hamiltonians $h$ of homogeneity-degree 2. This presumably relates to the deformed Poisson structure of \cite{Lipstein:2023pih}. The sigma-model of \cite{Adamo:2021bej} was already formulated also at $\Lambda\neq 0$ and its open disc version generalizes similarly to that given here.  This construction is being pursued further in  joint work with Claude Lebrun.

These computations are all done at the level of the classical, as opposed to quantum, sigma model.  This is in contradistinction to  the twistor-string or ambitwistor string which generate amplitudes via quantum correlators (and requires supersymmetry).  Clearly the twistor-strings of \cite{Witten:2003nn, Berkovits:2004jj, Skinner:2013xp}  and 4d ambitwistor strings of \cite{Geyer:2014fka} can  be similarly formulated as open string models, as above, even in terms of their gravity vertex operators. unlike conventional string theory.  Perhaps more interesting would be to formulate the higher-dimensional ambitwistor models of \cite{Mason:2013sva} in this way, and to understand the underpinning semi-classical geometry.

Following \cite{Adamo:2021bej} we remark that at MHV, the  sigma model correlator gives the theory whose Feynman tree-diagrams give precisely the  tree formulae of \cite{Bern:1998sv, Nguyen:2009jk}.  Furthermore, in that paper, it was shown that  the formula at MHV yields the Einstein-Hilbert action of a space-time generated by $n$ perturbations at $\scri$.  We expect a similar proof to be valid here too.

We now give some more extended discussion of further avenues for exploration.

\subsection{A one-loop all-plus generating function}
The ingredients introduced in in \cite{Adamo:2021bej} and reformulated above allow us to introduce a generating function for the all-plus 1-loop amplitude as originally presented in  \cite{Bern:1998xc,Bern:1998sv}.  
This was presented in terms of certain so called \emph{half-soft} factors $h(a,A,b)$  where the letters $a,b \in \{1,\ldots ,n\}$ 
and  $A\subset \{1,\ldots,n\}$.\footnote{The  $h(a,A,b)$ were defined to have half the soft limit expected for the gravity amplitude.}  The  $h(a,A,b)$ can be defined to be the sum over expressions corresponding to connected trees diagrams with vertices correponding to letters of $A$,  and for whom the propagators between vertices $i, j$ correspond to $\mathbb{H}_{ij}$ with $t_i=\langle i 1\rangle \langle i 2\rangle/\langle 1 2\rangle$, see \cite{Feng:2012sy} where it is also explained that the matrix-tree theorem can be used to give the definition
\begin{equation}
h(a,A,b)=\frac{1}{\prod_{i\in A} t_i^2}\det{}'\mathbb{H}_A
\end{equation}
where the reduced determinant $\det{}'\mathbb{H}_A$ is the determinant of a $|A|-1 \times |A|-1$ minor of $\mathbb{H}_{ij}$ with $i,j,\in A$. The final formula given by \cite{Bern:1998sv} is, up to an overall constant
\begin{equation}
\cM(1,\ldots,n)= \delta^4\left(\sum_i k_i\right)\sum_{\{a,A,b,B\}=\{1,\ldots, n\}} h(a,A,b)h(b,B,a)\label{1-loop} [a,A,b,B]^3\, ,
\end{equation}
where $[a,A,b,B]=\tr (\slash\!\!\! k_a\slash\!\!\! k_A\slash\!\!\! k_b\slash\!\!\! k_B)$, $k_A=\sum_{i\in A} k_i$ and the sum is over partitions $\{a,A,b,B\}$ of $\{1,\ldots, n\}$.

The main arguments of \S5 of \cite{Adamo:2021bej} include the proof that, at $k=2$ for the tree-level MHV amplitude,  the on-shell sigma model action $S^{os}(Z_i,\sigma_i)$, $i=1,2$, can be identified  with the first Plebanski scalar $\Omega(x,\lambda_1,\lambda_2)$  adapted to the two $\lambda_\alpha$-spinors of $Z_1$ and $Z_2$; from a Euclidean perspective, if $Z_1, Z_2$ are taken to be complex conjugate twistors, then $\Omega$ is the Kahler scalar in the complex structure determined by their $\lambda$-spinors. The  worldsheet M\"obius   coordinate freedom  removes the dependence  $S^{os}$ on the $\sigma_i$, $i=1,2$ and so $S^{os}$ depeofnds only on the point $x$ in  space-time that corresponds to the line in twistor space joining the two twistors $Z_1,Z_2$.
On the other hand, the subsequent part of \S5 in \cite{Adamo:2021bej}, as reprised in \S\ref{Open-model} above, expands the on-shell action, as a sum of tree diagrams giving rise to the formulae given in \cite{Bern:1998sv} for the tree-level MHV amplitude as a sum over trees.  The sigma model  therefore provides the theory that underpins that tree expansion. 
%discovered in \cite{Bern:1998sv} 
This is resummed
 into the reduced determinant formula above by \cite{Feng:2012sy}. With these identifications we see  that
\begin{equation}
h(1,A,2)\delta^4\left(\sum_{i\in\{1,2,A\}} k_i\right)=\int d^4x \, \e^{i(k_1+k_2)\cdot x} \left(\prod_{i\in A} \delta_i\right) \Omega(x, \kappa_1,\kappa_2)
\end{equation}
where $\delta_i$ denotes the variation with respect to the momentum eigenstate with momentum $k_i$.

We can now see that \eqref{1-loop} can be generated as the $n-2$th variation of the local two-point function on the self-dual background determined by $h$
\begin{equation}
\cM(1,2,h)= \int d^4x \, \left.\e^{i(k_1+k_2)\cdot x} \left[1,\p_{x},2,\p_{x'}\right]^3\Omega(x)\Omega(x')  \right|_{x=x'}\, ,
\end{equation}
in the sense that the $n-2$th variation of this formula restricted to flat space gives \eqref{1-loop} above. Both factors of $\Omega$ are the on-shell action of the sigma model with insertions at points $\kappa_1$ and $\kappa_2$, so together the two factors represent a loop, a bubble, on the worldsheet.  This is much the same as for the Yang-Mills 1-loop generating function presented in equation 
4.17 of \cite{Boels:2007gv}, which can, as written,  be thought of as a 2nd variation  associated to a bubble on the auxiliary WZW-model on the sphere.

\subsection{Yang-Mills and colour/kinematic duality}
The analogous case of global solutions to the self-dual Yang-Mills equations on $S^2\times S^2$ in split signature was treated in \cite{Mason:2005qu}.  Although there is now the possibility of introducing a nontrivial global holomorphic vector bundle on $\CP^3$, if we stick to  the topological sector containing  the trivial connection, the data of a self-dual Yang-Mills field on $S^2\times S^2$ is encoded in a hermitian matrix function $H(Z)$ of unit determinant on $\RP^3$ (up to some global constant frame change). The space of Hermitian matrices can be understood as the quotient of $SL(n,\C)/SU(n)$,  using the relation $H=GG^*$ analogous to  the quotient  $Lw_{1+\infty}^\C/Lw_{1+\infty}$.

The analogue to the chiral sigma model is to consider again the degree $k-1$ disk  $Z(\sigma)$  given in \eqref{marked-points} but now unperturbed
\begin{equation}
Z(\sigma)=\sum_{i=1}^k \frac{Z_i}{\sigma- \sigma_i}\, . \label{marked-points-YM}
\end{equation}
The analogue of finding  the deformed holomorphic discs, is now simply to   find the holomorphic  frame $F$ on the degree-$k-1$ discs above   with boundary condition given as
\begin{equation}
F:D\rightarrow SL(n,\C)\, , \qquad \dbar_{\bar\sigma} F=0, \qquad    FF^*|_{\p D}=H|_{\p D}\, .
\end{equation}
The reconstruction of the  connection on space-time is realized using the $F$ in found in the  $k=2$ case (i.e., for degree-1).
The analogue of proposition \ref{background-field-gamps} is now given by a formula for the amplitude for $k$ ASD gluons  on a self-dual Yang-Mills background.  If the $k$ ASD gluons are represented by the weight $-4$ functions $\tilde f_i$ on $\RP^3$, then the formula, with a choice of cyclic ordering,  is
\begin{equation}
\cA(H, \tilde h_i)= \int_{(\RP^3\times \RP^1)^k} %S_{WZW}^{os}(Z_i,\sigma_i) 
\prod_{i=1}^k F_i\tilde f_i(Z_i) F^*_i D^3Z_i \frac{d\sigma_i}{(\sigma_i\sigma_{i+1})}
\end{equation}
where %now $S_{WZW}^{os}(Z_i,\sigma_i)$ is the on-shell Wess-Zumino-Witten action of $F$ on $D$, and\footnote{This  makes sense, despite the apparent singularity of $Z(\sigma)$ above, as the fact that $F$ has homogeneity degree 0 means it doesnt change when multiplying $Z(\sigma)$ by $\prod_i (\sigma\sigma_i)$.} 
$F_i=F(Z_i)$.  Writing $H(Z)$  as a sum of perturbations  $H(Z)=\sum_{i=k+1}^n f_i$ and expanding $F(Z)$ using the standard formula for the variation of a Greens function, here on the disc, we obtain
\begin{equation}
G_{ij}:=\frac{ F^*_i(Z(\sigma_i)) F_j(Z(\sigma_j))}{\sigma_i-\sigma_j}\, ,\qquad \delta_k  G_{ij}=\int_\D  G_{ik} f_k G_{kj} d\sigma_k\, ,
\end{equation}
and using this $n-k$ times generates the full Parke-Taylor factor and  gives  the original RSVW formulae \cite{Roiban:2004yf}.   An action for this problem can be found following  the ideas of Nair \cite{Abe:2004ep,Nair:2005wh}, but now framed in terms of the boundary value problem for $F$ following from an open  Wess-Zumino action.

In comparing the Yang-Mills version to the gravitational one, the translation from colour to kinematics can be seen in the replacement of the Parke-Taylor factor by the  product of the two reduced determinants $\det{}'\mathbb{H} \det{}' \widetilde{\mathbb{H}}$ (as can also easily be seen by comparing the RSVW Yang-Mills formula to the Cachazo-Skinner gravity formula).    The framework of \cite{Adamo:2021lrv} and this paper shows that the
$\det{}'\mathbb{H} $ factor is associated with the $Lw_{1+\infty}$ algebra, and therefore $\det{}' \widetilde{\mathbb{H}}$ is associated with the algebra $\widetilde{Lw}_{1+\infty}$ that acts on dual twistor space.  The latter trivializes in the MHV sector hence explaining why $Lw_{1+\infty}$ is sufficient in  the MHV sector as described in \cite{Monteiro:2011pc}.  However, at arbitrary MHV degree, we will need both, but they will not commute or close onto something simple  classically as remarked in \cite{Adamo:2021lrv}, although  see \cite{Adamo:2021zpw} for how they can work together in the ambitwistor string.  This is clearly a further avenue for development.

\subsection{Continuation to Lorentz signature and encoding the full space-time data}

%Because in split signature $\scri$ only has one component, one does not aim to define a scattering matrix in terms of evolution from $\scri^-$ to $\scri^+$ as such.  One instead aims to compute the path integral over metrics with given boundary data at $\scri$. This should then, by analytic continuation to Lorentz signature, yield what is normally meant by a scattering amplitude. See \cite{Witten:2001kn, Atanasov:2021oyu} for related discussions. Nevertheless, the tree-level S-matrix itself, by crossing symmetry and analyticity, is valid in all signatures so that in particular the formulae continue to physical Lorentz signature.  

A subtlety of split signature is that the asymptotic shear $\bigma$ cannot be regarded as free data. As seen above, in linear theory it is even under the antipodal map.  In general, when fields are nonlinear, we expect that identification to become some nonlinear relation forced by the requirement that the intersection between $\beta$-planes and  $\scri$ are circles.  The split signature $\scri$ is the continuation of $\scri^+\cup\scri^-$ from Lorentz signature where again, we understand the shear on $\scri^+$ to be determined by that on $\scri^-$ or vice-versa (and they are again identified in linear theory). This analogy between the two frameworks was used in \cite{Mason:2005qu} to obtain a split-signature analogue of classical scattering in split signature  by dividing $\scri$ into two parts interchanged by the antipodal map; the relationship between the asymptotic data and its antipode required the reversal of ordering of the Riemann-Hilbert problem leading to nontrivial scattering even within the self-dual sector.  This is however  a nonlinear map that is no longer so trivial as to tell us that the data is even or odd under the antipodal map.

In Lorentz signature, 
%the global structure is of course quite different, with two disconnected components to $\scri$ interchanged in flat space by the antipodal $\Z_2$ map that played  a role throughout.  We then also 
we have the two distinct twistor spaces at $\scri^\pm$, each encoding the respective asymptotic shears.  One can alternatively speculate as to some gluing  arising 
 from the generating function that defines the scattering along null geodesics  from $\scri^-$ to $\scri^+$ near $i^0$ in a similar spirit to that   first introduced by Penrose in \cite{Penrose:1972ia}. From this perspective there is an ambiguity between the interpretation of $h$ as a patching between different sets covering one asymptotic twistor space, or between that at $\scri^-$ and at $\scri^+$; this latter view was that taken also in \cite{Geyer:2014fka}. From this perspective it is the non-triviality of the antipodal map that in Lorentz signature interchanges $\scri^\pm$ that is encoding the patching that generates  the $Lw_{1+\infty}$ motion that deforms the twistor space.  %However, in Lorentz signature, there are also deformations from the shear that provide deformations on those spaces defined only on $\scri^\pm$ too.  
In particular, the asymptotic shear misses out a key part of the gravitational data associated to the value at $i^0$ of the mass aspect, and perhaps more---these are `constants of integration' in the solution to the characteristic intiial value problem from $\scri$.  The fact that the shear at $\scri^-$ and that at $\scri^+$ are in some nonlinear relation that should also include the missing $i^0$ mass-aspect data makes the twistor-encodng of scattering incomplete in the nonlinear regime, but one can hope that there is some completion. 
 In ambitwistor space, it should possible to see these all as contributions to the patching.

\acknowledgments
It is a pleasure to acknowledge extended discussions with Claude LeBrun, Julio Parra Martinez,  Atul Sharma and David Skinner, and particularly Atul Sharma for also  providing feedback on the draft, together with the anonymous referee.   The author would also like to thank the IHES and ENS in Paris  for hospitality while this was being written up and the STFC for financial support from  grant number ST/T000864/1.  The present research was partially supported by the  `2021 Balzan Prize for Gravitation: Physical and
Astrophysical Aspects', awarded to Thibault Damour.
\\

\noindent 
{\bf Conflicts of interest and data availability:} There is no conflict of interest nor additional data available for this article.  This article appeared as eprint arxiv:2212.10895.

 \appendix

\section{Review of split signature integral formulae for massless fields.}\label{int-formulae}
Massless fields  of helicity $n/2$ are  given respectively  by  the formulae
\begin{align}
\Phi_{\alpha_1\ldots \alpha_n}(x)&=\int \kappa_{\alpha_1}\ldots\kappa_{\alpha_n} \hat \phi(\kappa,\tilde\kappa) \e^{ik\cdot x} \delta(k^2) d^4k \,,\\ 
&=\int  \lambda_{\alpha_1}\ldots \lambda_{\alpha_n} \p_u\phi^0( x^{\alpha\dot\alpha}\lambda_\alpha \tilde \lambda_{\dot\alpha},\lambda,\tilde \lambda) \langle \lambda d \lambda\rangle [\tilde \lambda d\tilde \lambda]\, ,\\
&=\int\lambda_{\alpha_1}\ldots \lambda_{\alpha_n} f(\lambda, x^{\alpha\dot\alpha}\lambda_\alpha) \langle \lambda d \lambda\rangle \, . \label{X-ray}
\end{align}
The first of these is the standard Fourier transform in split signature, the second is an adaptation of the D'Adhemar integral formula of, for example, \cite{Penrose:1984uia}, to split signature. The third is the so-called  \emph{X-ray transform}, because it integrates $f$ over lines in $\RP^3$,  first observed by Fritz John as providing solutions to the ultrahyperbolic scalar wave equation, \cite{John:1938}.\footnote{Usually, the Penrose transform reformulates massless fields as cohomology classes on regions in twistor space and as such have constraints and gauge freedom.  The X-ray transform has the advantage of fixing this.} This latter is conformally invariant.

Concatenating these leads to Witten's \cite{Witten:2003nn} \emph{half-Fourier} transform between momentum space and twistor space
\begin{equation}
f(\lambda, \mu)=\frac{1}{\sqrt{2\pi}}\int \hat \phi(\lambda,\tilde \lambda) \e^{i [\mu\, \tilde \lambda]} d^2\tilde \lambda\, , \qquad \hat\phi(\lambda,\tilde \lambda)= \frac1{\sqrt{2\pi}}\int f (\lambda,\mu) \e^{-i[\mu\, \tilde{\lambda}]} d^2\mu\,, \label{half-fourier}
\end{equation}
see for example appendix B of \cite{Mason:2009sa} for    details.  Similarly we can use the conformal invariance of the X-ray transform to give the asymptotic data in terms of $f$ as 
\begin{equation}
\phi^0(u, \lambda,\tilde \lambda)=\int_{-\infty}^\infty f(\lambda_\alpha, \mu^{\dot \alpha}+ t \tilde \lambda^{\dot\alpha})\, d t\, ,\label{Radon}
\end{equation}
where $u=[\mu\,\tilde \lambda]$.  
This is essentially a Radon transform in the $\lambda_\alpha=$ constant plane in which one integrates $f$ over lines in that plane.  This is related to the so-called Funk transform in which one takes a double cover of the projective plane to obtain the 2-sphere, $S^2$ and integrates over great circles (and therefore has kernel consisting of functions that are odd under the antipodal map).   The inversion of this is more non-local, a two-dimensional integral, see for example \cite{Bailey:1999, Bailey:2002} and is related to the light-ray transform \cite{Sharma:2021gcz}.   

Combining this with the half-Fourier transform, after integrating out $t$ to obtain a delta-function, we obtain what might be known as the \emph{third-Fourier transform}:
\begin{equation}
\phi^0(u,\lambda,\tilde{\lambda})=\int \hat\phi(\lambda, s\tilde\lambda) \e^{isu} ds\, .
\end{equation}

We finally note the positive homogeneity versions of equations \eqref{X-ray}  
\begin{equation}
\Phi_{\dot \alpha_1\ldots \dot\alpha_n}(x)=\int\left.\frac{\p^n  f}{\p\mu^{\dot\alpha_1}\ldots\p\mu^{\dot\alpha_n}}\right|_{\mu^{\dot\alpha}=x^{\alpha\dot\alpha}\lambda_\alpha} \langle \lambda d \lambda\rangle \, ,
\end{equation}
and \eqref{Radon}
\begin{equation}
\phi^0(u, \lambda,\tilde \lambda)=\int_{-\infty}^\infty \frac{d^n}{d t^n} f(\lambda_\alpha, \mu^{\dot \alpha}+ t \tilde \lambda^{\dot\alpha})\, d t\, ,
\end{equation}

In the case of linear gravity the
we denote the linear Weyl spinor by $\psi_{\dot\alpha\dot\beta\dot\gamma\dot\delta}$ and the twistor function of homogeneity degree 2 by $h(Z)$ for which  the higher spin extension of the X-ray transform is given by
\begin{equation}
\psi_{\dot\alpha\dot\beta\dot\gamma\dot\delta}(x)= \oint_{\mu^{\dot\alpha}=x^{\alpha\dot\alpha}\lambda_\alpha} \frac{\p^4 h}{\p\mu^{\dot\alpha}\p\mu^{\dot\beta}\p\mu^{\dot\gamma}\p\mu^{\dot\delta}} D\lambda\, , \qquad D\lambda=\langle\lambda\, d\lambda\rangle\, 
.\label{X-ray-plus}
\end{equation}
The corresponding characteristic data at $\scri$ is usually denoted by $\psi^0_4$ and we have 
$$\psi^0_4=\p_uN=\p^2_u \sigma
$$ 
where $N$ is the news, and $\sigma$ the asymptotic shear.  For the $\scri$ quantities  we have the integral formulae that follow from
\begin{equation}
\bigma=\p^2_u\int_{-\infty}^\infty  h(\lambda_\alpha, \mu^{\dot \alpha}+ t \tilde \lambda^{\dot\alpha})\, d t\, .\label{scri-radon}
\end{equation}
This formula is inverted at least in part by the light ray transform in \cite{Strominger:2021lvk,Sharma:2021gcz}.

In the following we will need to insert twistor functions that give rise to momentum eigenstates and these are given uniformly by
\begin{equation}
f(Z)=\int \frac{dt}{t^{1+n}} \delta^2(t\lambda_\alpha -\kappa_\alpha)\e^{i t [\mu,\tilde\kappa]}
\end{equation}
where the parameter $t$ is integrated against the delta function to yield a function of homogeneity degree $n$ in $Z$.

We now turn to describe the nonlinear analogues of these transforms based on Penrose's nonlinear graviton construction.

\section{Examples with symmetry and  SD black holes}\label{taub-nut}
The Gibbons-Hawking ansatze has an analytic continuation to split signature that can be described as follows. These are solutions with a symmetry that preserves the pseudo-HyperKahler structure. Our symmetry is defined by a nondegenerate 4-vector  $T^{\alpha\dot\alpha}$ of squared length 2.  This can be used also to convert all dotted indices to undotted, i.e., $\mu^\alpha:=T_{\dot\alpha}^\alpha\mu^{\dot\alpha}=u^{\alpha}+iv^{\alpha}$ etc..  The generating function $h$ will be invariant if we take it of the form 
$$
h=h(\mu^\alpha\lambda_\alpha,\lambda_\alpha)\, .
$$ 
The deformed real slice $\PT_\R$   is then defined by
$$
v^\alpha = \lambda^\alpha \dot h\, .
$$

In order to find the holomorphic discs with boundary in this $\PT_\R$,  use $\lambda_\alpha$ as homogeneous coordinates on the holomorphic   disks so that its components can be taken to be real on the boundary.  Then the discs can be  expressed as sections of the fibration to $\D$  by
$$
\mu^\alpha=x^{\alpha\beta}\lambda_\beta + (t+g(x,\lambda)) \lambda^\alpha, \qquad x^{\alpha\beta}=x^{(\alpha\beta)}\, .
$$
 where 
\begin{equation}
g(x^{\alpha\beta},\lambda)=\frac1{2\pi}\oint \frac{\lambda_0}{\lambda'_0}\frac{1}{\langle\lambda  \lambda'\rangle} \dot h((x^{\alpha\beta}\lambda'_\alpha\lambda'_\beta,\lambda'_\alpha) D\lambda'\, ,
\end{equation}
with contour $\p \D$.  This follows because on taking the  imaginary part, the contour can be deformed to one around $\lambda=\lambda'$ so that
\begin{equation}
\Im g(x^{\alpha\beta},\lambda)=  \dot h((x^{\alpha\beta}\lambda_\alpha\lambda_\beta,\lambda_\alpha) \, ,
\end{equation}
by the Cauchy residue theorem,  as required.

Working through the construction  gives split signature version of Gibbons-Hawking metrics
\begin{equation}
ds^2 =V^{-1} (dt+\omega)^2+ Vd \bx\cdot d\bx  \, ,\label{Gibbons-Hawking}
\end{equation}
where $d \bx\cdot d\bx$ denotes the signature $1+2$-dimensional flat metric and
\begin{equation}
 d V=^*d\omega\, , \qquad V=1+\oint\ddot{h}D\lambda\, ,\qquad \lambda^\beta\p_{\alpha\beta} g=\lambda^\beta\omega_{\alpha\beta}+\lambda_\alpha V\, .
\end{equation} 
As a consequence, $V$ satisfies the $1+2$-dimensional wave equation.  By construction, 
the metric has a Killing vector $\p_t$:
note that in general, this Killing vector  obstructs asymptotic flatness in that direction.  However, for small data, these metrics are  
%Asymptotically, we can write $d \bx\cdot d\bx= dz^2- d\rho^2-\rho^2d\theta^2$ and assuming that $V\sim 1+ O(1/\rho)$ and $\omega\sim O(1/\rho)$ for large $\rho$ we can write
%\begin{equation}
%ds^2=ds_0^2+\frac{1}{\rho}
%\end{equation}

A special case that can be constructed by this ansatze  is the split signature analogue of the self-dual Schwarzschild/Taub-Nut, i.e., the metric whose mass is equal to its NUT charge. This can no longer be considered as `small'.  Its analytic continuation to split signature was described in \cite{, Crawley:2021auj}.  It is  singular, so strictly speaking does not satisfy the global conditions of this paper.   To obtain this case,  take $h=Q\log Q$ where $Q$ is the quadratic form  defined by $Q=\lambda_\alpha\mu^\alpha$.  This is clearly singular on $\RP^3$, and the  integral formulae above for $V$ needs some $i\epsilon$  prescription to make sense  to give 
\begin{equation}
V=1+\frac{2m}{r}
\end{equation}
and we can write the split signature metric \eqref{Gibbons-Hawking} in the form
\begin{equation}
ds^2=\frac{1}{4}\left(1+\frac{2m}{r}\right)( dr^2 - r^2 d\Omega^2) + m^2 \frac{1}{1+\frac{2m}r}\sigma_3^2\, .
\end{equation}
Here we have expressed the flat $1+2$-dimensional flat metric as 
\begin{equation}
ds^2_{1+2}:=dr^2 - r^2 d\Omega^2=dr^2-r^2(d\theta^2+\sinh^2\theta d\phi^2)\, ,
\end{equation}
and 
 \begin{equation}
\sigma_3 = dt-\cosh \theta d\phi\, .
 \end{equation}
As observed by \cite{Crawley:2021auj}, these are natural Wick rotations of the metric on $S^3=SU(2)$ to AdS$_3 = SU(1,1)$ in which $d\Omega^2$ is continued from the round sphere to the hyperbolic disc, and $\sigma_3$, the left-invariant form dual to the generator  $\sigma_3$ in $SU(2)$, is continued to the corresponding form in $SU(1,1)$.
% This  gives the split signature version of the SD %Taub-NUT metric
%\begin{equation}
%ds^2=\frac{1}{4}\left(1+\frac{2m}{r}\right)ds^2_{1+2}  + m^2 \frac{1}{1+\frac{2m}r}\sigma_3^2\, .
%\end{equation}
%In this metric we have the tori parametrized by $(t,\phi)$ and so can draw 2d diagrams in the $(r,\theta)$ plane as in \cite{Crawley:2021auj}.
The metric above converts to the expression 
%\begin{equation}
%ds^2=m^2 \frac{r-m}{r+m}\sigma_3^2+ \frac{1}{4}\frac{r+m}{r-m} dr^2 -\frac{1}{4}(r^2-m^2) d\Omega^2\, .
%\end{equation}
of \cite{ Crawley:2021auj}  by shifting $r-m\rightarrow r$.

\bibliography{sdpw1}
\bibliographystyle{JHEP}

\end{document}